\documentclass[11pt]{article} \usepackage[preprint]{mldraft}
\usepackage{xcolor}

\definecolor{MyGreen}{HTML}{009900}
\RequirePackage[colorlinks,citecolor=MyGreen]{hyperref}
\usepackage{url}
\usepackage{enumitem}
\newcommand{\model}{{Fusion-pMT}}

\newcommand{\mtx}[1]{\bm{#1}}

\newcommand{\defeq}{\vcentcolon=}

\title{\model: Biological Sequence Modeling for \\Tri-Molecular Binding in Immunogenicity Prediction}

\author{
\centerline{
\name Jiahao Ma$^{1,3}$
~~
\name Yujia Yin$^{2}$
~~
\name Hongzong Li$^3$
~~
\name Yunke Shi$^{2}$
~~ 
\name Yifan Wu$^{2}$
} \\[1ex]
\centerline{
\name Jian-Dong Huang$^{1}$
~
\name Ye-Fan Hu$^{3}$
~
\name Yifan Chen$^{2}$
} \\[1ex]
\centerline{$^1$ \addr Li Ka Shing Faculty of Medicine, The University of Hong Kong}
\centerline{$^2$ \addr Department of Computer Science, Hong Kong Baptist University}
\centerline{$^{3}$ \addr BayVax Biotech Limited}
}

\usepackage{latexsym}
\usepackage{colortbl}
\usepackage{microtype}
\usepackage{graphicx}
\usepackage{url}
\usepackage{booktabs}
\usepackage{amsfonts}
\usepackage{amssymb}
\usepackage{mathtools}
\usepackage{bm}
\usepackage{nicefrac}
\usepackage{multirow}
\usepackage{subcaption}
\usepackage{placeins}
\usepackage{float}
\usepackage{enumitem}
\usepackage{tikz}
\usepackage{soul}
\usepackage{pifont}
\usepackage{bigstrut}
\usepackage[capitalize,noabbrev]{cleveref}
\makeatletter
\AddToHook{cmd/appendix/before}{\def\cref@section@alias{appendix}}
\makeatother

\usepackage{wrapfig}

\usepackage{makecell}

\begin{document}

\maketitle
\lhead{}

\begin{abstract}
Accurately modeling the tripartite interaction among peptides, major histocompatibility complex (MHC) molecules, and T cell receptors (TCRs) is essential for understanding T-cell-mediated immunity and informing immunogenicity prediction. A key challenge is to connect peptide--MHC binding knowledge with TCR specificity while retaining the residue-level context needed to model their interactions. In this paper, we propose Fusion-pMT, a biological sequence modeling framework that (1) learns representations of the three molecular inputs through a shared amino acid projection and molecule-specific encoders, and (2) combines peptide--MHC pretraining with two-stage cross-attention, retaining MHC-conditioned peptide tokens for subsequent interaction with TCR sequences. By aligning sequence fusion with the stepwise nature of immune recognition, Fusion-pMT effectively supports both pairwise and tripartite interaction modeling. To validate its effectiveness, we conduct comprehensive experiments covering both pairwise and tripartite interactions and demonstrate the advantages of Fusion-pMT over strong baselines while validating its core design choices.

\end{abstract}

\section{Introduction}

T cell recognition relies on an antigenic peptide presented by a major histocompatibility complex (MHC) and a T cell receptor (TCR) engaging the resulting pMHC complex~\citep{Huppa2010pMHCTCR}. Accurately predicting peptide--MHC--TCR binding is critical for advancing cancer vaccines and TCR therapies~\citep{Rojas2023NeoAgtrial,Haseel2023TCRTtrial,Angelo2024TCRT}, while peptide--MHC binding affinity itself substantially influences tumor recognition~\citep{Kammertoens2013pMHC}. However, annotated three-dimensional structures of these complexes remain scarce, making structure-based docking predictions computationally challenging~\citep{bradley2023structure}. To bypass this limitation, sequence-based machine learning methods offer a highly scalable approach to learn binding patterns directly from primary amino acid sequences.

Existing sequence-based triad predictors differ in how they represent the intermediate peptide--MHC complex and propagate this information to the TCR interaction. For instance, pMTnet~\citep{lu2021pMTnet} pretrains separate modules for peptide--MHC and TCR inputs but compresses each into a fixed-dimensional vector before concatenation, rather than explicitly computing residue-level cross-attention between the sequences. Conversely, PISTE~\citep{feng2024sliding} models residue-level interactions across all three sequences via a sliding-attention transformer; however, it does not incorporate a mechanism to transfer a separately pretrained peptide--MHC binding stage into its tripartite framework.

\begin{figure*}[t!]
  \centering
  \includegraphics[width=\textwidth,keepaspectratio]
  {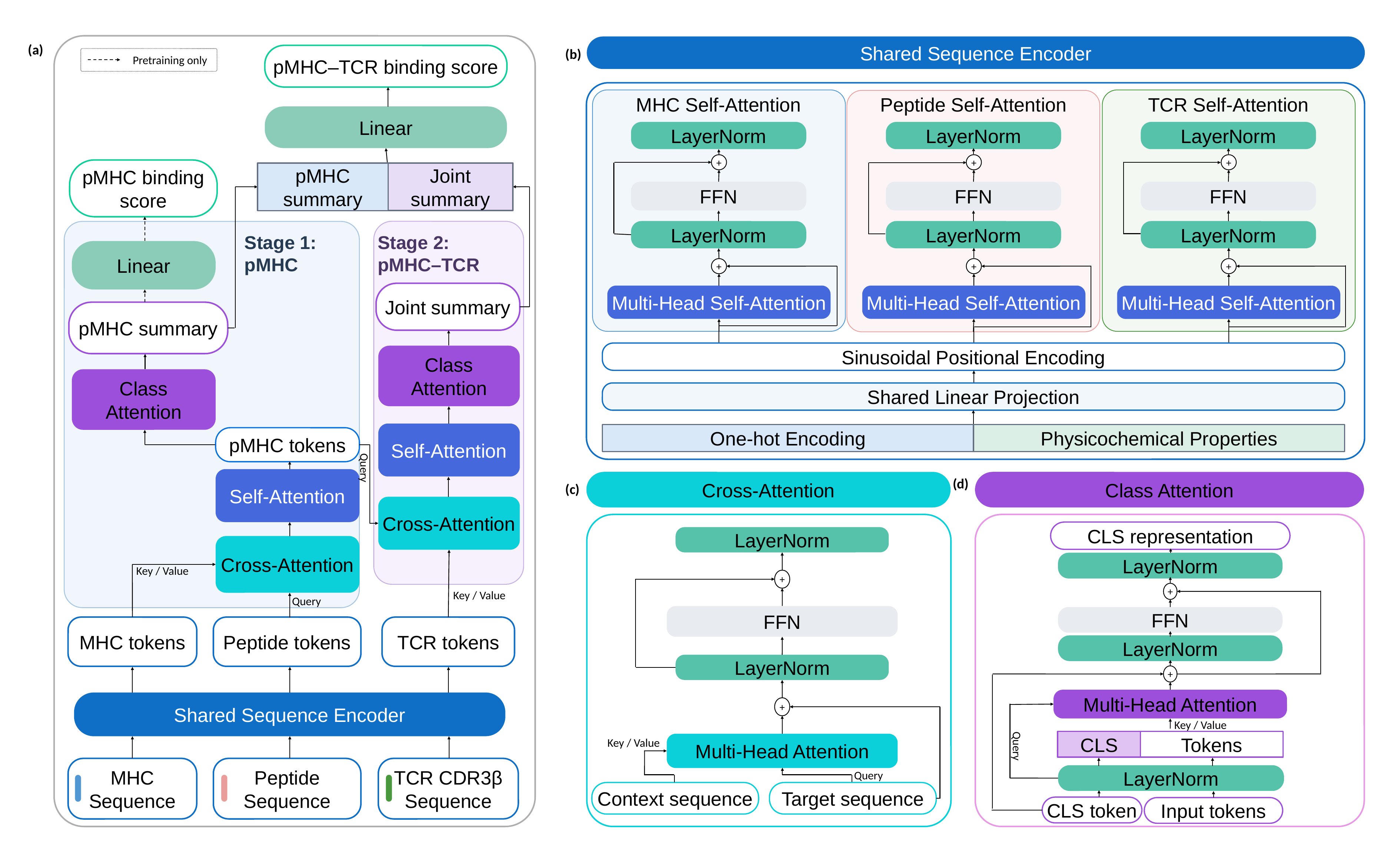}
\caption{\textbf{(a)} \textbf{The overall Fusion-pMT framework}. Three molecular sequences are processed by the \textbf{(b)}~\textbf{shared sequence encoder}. 
{The peptide, MHC, and TCR streams share a linear projection (embedding) while using independently parameterized self-attention and feed-forward blocks.} 
In Stage~1 (\textit{i.e.}, Fusion-pM), we employ \textbf{(c)}~\textbf{cross-attention} to form MHC-conditioned peptide tokens and \textbf{(d)} \textbf{class attention} to generate a pMHC summary for binding prediction. These retained tokens then query the TCR tokens in Stage~2, transferring pairwise binding knowledge into the receptor-specific fusion. A second class-attention readout yields a joint summary, which is combined with the pMHC summary for the final tripartite prediction. Dashed connections indicate the peptide--MHC prediction branch used solely during Stage~1 pretraining.}
\label{fig:Model_Structure}
\end{figure*}

Building on these design choices, Fusion-pMT integrates pairwise peptide--MHC pretraining with a residue-level two-stage fusion architecture. A shared bias-free input projection maps residue features into a common latent space, followed by molecule-specific self-attention and feed-forward blocks. Stage 1 (Fusion-pM) is optimized on peptide--MHC binding data to generate a refined, position-specific pMHC sequence representation. Instead of collapsing the pMHC complex into a single vector, Stage 2 leverages this sequence representation as queries over TCR keys and values to preserve fine-grained residue-level context. A class-attention readout in Stage 1 summarizes the pMHC tokens for standalone peptide--MHC binding prediction. The final classifier synergizes outputs from both stages for comprehensive tripartite prediction (\Cref{fig:Model_Structure}). 
Our main contributions are:

\begin{itemize}[
topsep=0.0ex,
itemsep=0.5ex,
leftmargin=*
]
\item We develop Fusion-pMT, a two-stage peptide--MHC--TCR predictor that propagates residue-level pMHC tokens into a dedicated cross-attention stage, enabling direct, fine-grained interaction with TCR tokens.
    \item We introduce a standalone peptide--MHC binding predictor in Stage 1 that achieves robust pairwise prediction while seamlessly initializing the corresponding modules of the complete tripartite model.
    \item We comprehensively evaluate both prediction tasks against retrained baselines and public checkpoints, accompanied by extensive ablation studies that validate our architectural, training, and representational design choices.
\end{itemize}

\section{Related Works}

The prediction of interactions among peptides, MHC, and TCR is crucial in immunoinformatics.
For sequence-based prediction, the central modeling choices are how to represent each molecule and how to combine the resulting representations across interaction partners.
We review related methods by sequence modeling and interaction mechanisms. A broader comparison is provided in \Cref{tab:related}.

\textbf{Immunological sequence modeling.}
Historically, sequence representations typically relied on position-specific scoring matrices (PSSMs) to predict peptide–MHC binding~\citep{parker1994PSSM}. Sequence encoders learn the molecular context that subsequent interaction models use. 
For instance, NetTCR~\citep{montemurro2021nettcr} employed CNN encoders for TCR and antigenic peptides, and models such as DeepAttentionPan~\citep{jin2021DeepAttentionPan} and CapsNet-MHC~\citep{kalemati2023capsnet} enhanced traditional CNNs with attention layers to improve feature extraction.
Attention-based encoders extend this contextual modeling across sequence positions.
TransPHLA~\citep{chu2021transmut,chu2022transphla} incorporated self-attention modules to capture complex dependencies.
STMHCpan~\citep{ye2023stmhcpan} encodes peptide sequences and 34-residue MHC class I pseudosequences and uses a Star-Transformer architecture to predict peptide--MHC class I binding.
These advances improve the representations available for binding prediction, while the interaction module determines how information passes between molecules.

\textbf{Molecular binding mechanisms.}
Prior models employ diverse fusion strategies to capture pairwise interactions: NetTCR~\citep{montemurro2021nettcr} relies on the direct concatenation of hidden embeddings, while TransPHLA~\citep{chu2022transphla} leverages self-attention. Building on attention mechanisms, UniTCR~\citep{gao2024unitcr} incorporates cross-attention between TCR and epitope representations using a TCR encoder pretrained with paired gene-expression data. 

Fundamentally, these fusion choices determine whether the prediction relies on a joint sequence context, a relation between pooled features, or directed attention between residue representations. While these approaches are effective for pairwise tasks, tripartite prediction introduces a distinct biological constraint: TCR recognition strictly depends on the peptide within its presenting MHC context. Consequently, a tripartite model must effectively connect this joint peptide--MHC information with the TCR representation to accurately predict recognition.

\textbf{Models for Peptide--MHC--TCR Triad Binding.}
Existing triad models address this connection through varying architectural choices. 
In particular, pMTnet~\citep{lu2021pMTnet} encodes peptide--MHC pairs and TCR sequences using separately pretrained modules (an LSTM-based reimplementation of NetMHCpan and an autoencoder, respectively), and concatenates them to model the interaction. 

While this modular design transfers pairwise knowledge through fixed-dimensional summaries, it leaves residue-specific pMHC--TCR interactions implicit in the final fusion step. Conversely, PISTE computes sequential peptide--MHC and pMHC--TCR sliding-attention phases over residue representations, but lacks a separate peptide--MHC pretraining stage~\citep{feng2024sliding}. UniPMT jointly learns these relationships through a heterogeneous graph and multitask training~\citep{zhao2025unipmt}. However, bridging separately supervised peptide--MHC pretraining with explicit residue-level TCR fusion requires an intermediate sequence representation that can serve both stages. Fusion-pMT provides this interface by retaining the MHC-conditioned peptide tokens produced by its pretrained peptide--MHC modules. These tokens grant the subsequent attention stage direct access to position-specific pairwise binding features, while a separate readout supports independent peptide--MHC prediction.
{\Cref{tab:related-triad} summarizes these design choices for pMTnet, PISTE, and Fusion-pMT, which all involve sequence modeling.}

\begin{table}[t]
    \centering
    \footnotesize
    \caption{{Comparison of pMHC representations, TCR fusion, and pMHC training strategies in three tripartite binding models.}}
    \label{tab:related-triad}
    \resizebox{\textwidth}{!}{
    \begin{tabular}{@{}lp{0.25\linewidth}p{0.26\linewidth}p{0.32\linewidth}@{}}
        \toprule
        \textbf{Model} & \textbf{pMHC representation} & \textbf{Fusion with TCR} & \textbf{pMHC training and transfer} \\
        \midrule
        pMTnet & A single pMHC vector & Concatenates pMHC and TCR vectors & Pretrains the pMHC encoder for feature extraction \\
        PISTE & Residue-level pMHC features & Applies sequential peptide--MHC and pMHC--TCR sliding attention & No separate peptide--MHC pretraining stage \\
        \textbf{Fusion-pMT} & Residue-level pMHC features at peptide positions & Uses these position-specific features as queries over TCR tokens & Pretrains the pMHC modules, then fine-tunes them for tripartite prediction \\
        \bottomrule
    \end{tabular}
    }
\end{table}

\textbf{Incorporating Structural Information.}
While structural modeling provides a complementary description of molecular recognition, it relies heavily on available structural priors. For instance, \citet{bradley2023structure} used AlphaFold~2~\citep{jumper2021af2} to predict pMHC--TCR complex structures, but required known docking templates to guide the folding process. Their evaluation highlights that structure-based prediction is strictly bottlenecked by the extreme scarcity of annotated ternary complexes~\citep{bradley2023structure}. To circumvent this limitation, our study follows the mainstream paradigm by focusing exclusively on \emph{sequence-level modeling} across peptide, MHC, and TCR sequences, providing a highly scalable framework to prioritize biologically meaningful candidates for downstream immunological investigation and targeted experimental validation.

\section{Preliminaries}

This section introduces the biological and computational foundations of molecular recognition prediction. We first describe antigen presentation and recognition in \Cref{sec:immune-sys} and define the associated prediction tasks in \Cref{sec:immune-task}. \Cref{sec:ca-in-bio} then introduces residue sequence representations and attention operations for contextual encoding, sequence interaction, and prediction readout.

\subsection{Biological Sequences in Immune Systems}
\label{sec:immune-sys}

In computational immunology, peptides, MHC molecules, and TCRs are represented as ordered amino acid sequences, where their residue identities and sequence positions provide the primary inputs for learning binding specificity. Biologically, this specificity drives adaptive immunity through two sequential processes: \textbf{antigen presentation} and \textbf{antigen recognition}. First, antigen-presenting cells (APCs) bind antigenic peptides to MHC molecules to form peptide--MHC (pMHC) complexes. Next, T cells \textbf{recognize} these complexes via their T cell receptors (TCRs), establishing ternary pMHC--TCR complexes~\citep{Huppa2010pMHCTCR}. T cells harbor highly diverse TCRs composed of $\alpha$ and $\beta$ chains, with $\beta$-chain diversity playing a central role in distinguishing self from non-self antigens~\citep{Mora2019Diveristy}. 

\begin{figure}[t]
    \centering
    \includegraphics[width=0.57\linewidth]{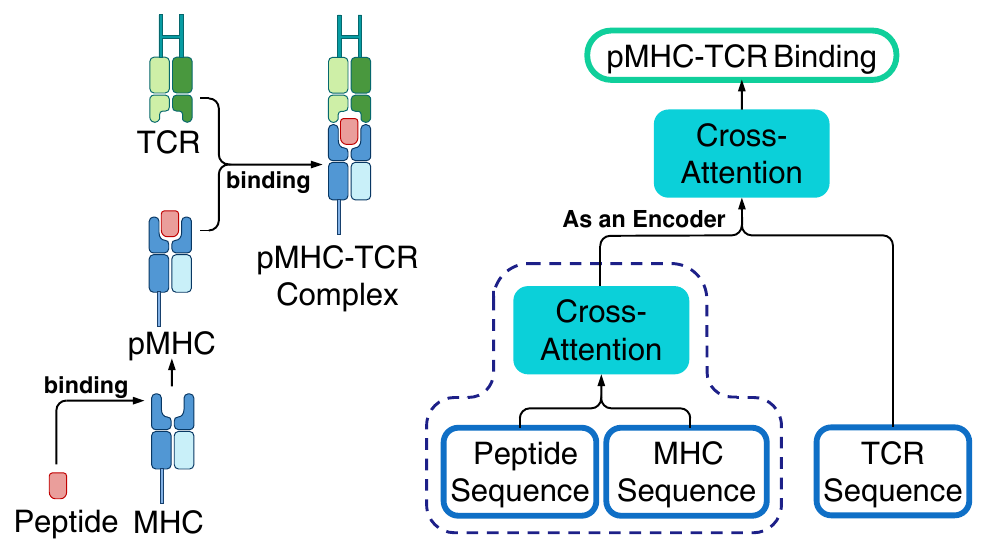}
    \caption{
    {\textbf{Antigen presentation and two-stage modeling.} The left panel shows peptide binding to MHC followed by TCR recognition. The right panel shows the corresponding Fusion-pMT workflow. The dashed outline marks the pretrained peptide--MHC module.}
    }
    \label{fig:T-cell}
\end{figure}

Because a successful immune response strictly requires both MHC-mediated presentation and TCR-mediated recognition, 
{\Cref{fig:T-cell} illustrates how Fusion-pMT maps these stepwise biological events directly to peptide--MHC pretraining and subsequent TCR fusion.} A broader overview of immune molecules is provided in \Cref{app:immune-bio-mol}.

\subsection{Molecular Binding Tasks}
\label{sec:immune-task}

Computational modeling of T cell recognition commonly considers three sequence-level formulations: \ding{182}~\textbf{Peptide--MHC binding}, which evaluates a prerequisite for antigen presentation (though binding alone does not guarantee TCR recognition); \ding{183}~\textbf{Peptide--TCR binding}, which omits the presenting MHC context and falls outside our scope; and \ding{184}~\textbf{Peptide--MHC--TCR interaction}, which jointly models all three sequences to capture full tri-molecular recognition specificity. Our framework addresses both task \ding{182} (via Stage~1, \textbf{Fusion-pM}) and task \ding{184} (via Stage~2, \textbf{Fusion-pMT}).

Given input sequences $\mtx X_p, \mtx X_M$, and $\mtx X_T$ (for the peptide, MHC, and TCR, respectively), we formulate the \textbf{peptide--MHC--TCR interaction prediction} task~\ding{184} as a binary classification problem over the interaction label $Y_{\mathrm{pMT}}\in\{0,1\}$:
\[
\Pr\!\left(Y_{\mathrm{pMT}}=1\mid\mtx X_p,\mtx X_M,\mtx X_T\right).
\]
Similarly, the \textbf{peptide--MHC binding prediction} task~\ding{182} models the binary label $Y_{\mathrm{pM}}\in\{0,1\}$:
\[
\Pr\!\left(Y_{\mathrm{pM}}=1\mid\mtx X_p,\mtx X_M\right).
\]
Both tasks are thus trained with binary cross-entropy loss using datasets comprising recorded positive interactions and constructed negative pairings. While the resulting predicted scores quantify interaction propensity under this labeling protocol, functional T cell activation ultimately depends on additional biological processes. An overview of the biological molecules involved is provided in Table~\ref{tab:antigen_presentation} of Appendix~\ref{app:immune-bio-mol}.

\subsection{Sequence Representations and Attention}
\label{sec:ca-in-bio}

\noindent\textbf{Residue representations.} For an input sequence $\mtx X\in\{\mtx X_p,\mtx X_M,\mtx X_T\}$ of $L$ residues, let 
{$\mathbf{F}\in\mathbb{R}^{L\times d_{\mathrm{in}}}$} contain its residue features. The initial sequence matrix then is
\begin{equation}
\label{eq:residue-representation}
\mathbf{H}^{(0)}=\mathbf{F}\mathbf{W}_{\mathrm{in}}+\mathbf{P}\in\mathbb{R}^{L\times d_{\mathrm{model}}},
\end{equation}
where $\mathbf{W}_{\mathrm{in}}\in\mathbb{R}^{d_{\mathrm{in}}\times d_{\mathrm{model}}}$ is a learned projection and $\mathbf{P}\in\mathbb{R}^{L\times d_{\mathrm{model}}}$ supplies positional information. Below, for convenience, $\mathbf{H}$ denotes the current sequence matrix, initially given by $\mathbf{H}^{(0)}$.

\noindent\textbf{Self-attention.} Self-attention derives queries, keys, and values from the same sequence~\citep{vaswani2017attention}. We formulate the single-head operations with a hidden dimension $d=d_{\mathrm{model}}/h$, where $h$ is the number of equal-width heads. 
{Each attention head uses learned projection matrices $\mathbf{W}^{Q},\mathbf{W}^{K},\mathbf{W}^{V}\in\mathbb{R}^{d_{\mathrm{model}}\times d}$. For notational simplicity, we omit head and module indices; each head in each attention module has its own parameters.} The operations are defined as:
\begin{equation}
\label{eq:self-attn}
\begin{aligned}
\operatorname{Attn}(\mathbf{Q},\mathbf{K},\mathbf{V})
&=\operatorname{softmax}\!\left(\frac{\mathbf{Q}\mathbf{K}^{\mathsf T}}{\sqrt d}\right)\mathbf{V},\\
\operatorname{SelfAttn}(\mathbf{H})
&=\operatorname{Attn}(\mathbf{H}\mathbf{W}^{Q},\mathbf{H}\mathbf{W}^{K},\mathbf{H}\mathbf{W}^{V}).
\end{aligned}
\end{equation}
Multi-head attention then concatenates these $h$ parallel outputs and linearly projects them back to the original dimension $d_{\mathrm{model}}$.

\noindent\textbf{Cross-attention.} By enabling one sequence to selectively attend to relevant parts of another, cross-attention has become a powerful mechanism for interaction modeling~\citep{hou2019cross, chen2021crossvit, ju2021joint, jin2023capla}. Its applications span diverse multimodal tasks, including text translation~\citep{gheini2021cross}, image captioning~\citep{zhang2023cross}, and voice recognition~\citep{sun2021multimodal}, as well as sequence-level predictions in immunology, such as modeling peptide--MHC and peptide--TCR specificity~\citep{yu2025unifyimmun}.

Given query sequence $\mathbf{H}_q\in\mathbb{R}^{L_q\times d_{\mathrm{model}}}$ and context sequence $\mathbf{H}_c\in\mathbb{R}^{L_c\times d_{\mathrm{model}}}$, cross-attention gives:
\begin{equation}
\label{eq:cross-attn}
\operatorname{CrossAttn}(\mathbf{H}_q,\mathbf{H}_c)
=\operatorname{Attn}(\mathbf{H}_q\mathbf{W}^{Q},\mathbf{H}_c\mathbf{W}^{K},\mathbf{H}_c\mathbf{W}^{V}).
\end{equation}
Crucially, because $\mathbf{K}$ and $\mathbf{V}$ originate from the same context sequence $\mathbf{H}_c$~\citep{chen-etal-2022-empowering}, the output effectively aligns external biological context (e.g., from MHC or TCR sequences) while strictly retaining the original $L_q$ query positions.

\noindent\textbf{Class-attention readout.} To aggregate sequence features for final prediction, a trainable classification (CLS) token $\mathbf{c}\in\mathbb{R}^{1\times d_{\mathrm{model}}}$ serves as a single query~\citep{touvron2021going}. By concatenating this token with the input sequence to form $\mathbf{Z}=[\mathbf{c};\mathbf{H}]$, the attention aggregation is formulated as:
\begin{equation}
\label{eq:class-attn}
\operatorname{ClassAttn}(\mathbf{c},\mathbf{H})
=\operatorname{Attn}(\mathbf{c}\mathbf{W}^{Q},\mathbf{Z}\mathbf{W}^{K},\mathbf{Z}\mathbf{W}^{V}).
\end{equation}
This single-query output updates the CLS representation to yield a prediction summary, while the sequence tokens remain untouched and available for subsequent interactions. Residual updates and normalization details for the readout are specified in \Cref{sec:readout-prediction,sec:imple-details}.

\section{Remodeling Peptide--MHC--TCR Triad Binding as Sequence Fusion}
\label{sec:remodeling}

In this section, we introduce \emph{Fusion-pMT}, a two-stage model that employs a refined design for sequence fusion, centered on a key premise: TCR recognition depends on the MHC context in which a peptide is presented.
We first construct contextualized residue sequences using a shared input projection and molecule-specific encoders (\Cref{sec:seq-rep}). The MHC-conditioned peptide sequence then provides the interface between peptide--MHC pretraining and TCR interaction modeling (\Cref{sec:dissection}). 
{Lastly, \Cref{sec:readout-prediction} describes the class-attention readouts and prediction heads.}

\subsection{Residue Representation and Contextual Encoding}
\label{sec:seq-rep}
\label{sec:unified-enc}

To model molecular interactions effectively, the framework must first translate raw amino acid strings into contextualized continuous representations. We achieve this by combining a shared vocabulary space with molecule-specific contextual encoders.

\noindent\textbf{Input preparation.} To accommodate the naturally variable lengths of biological sequences, inputs are padded to a uniform tensor width, with these padding positions explicitly masked in all attention and class-attention readouts to prevent them from influencing the learned representations. MHC molecules are represented using the standard 34-residue pseudosequences.

\noindent\textbf{Residue features and shared projection.} Because the peptide, MHC, and TCR molecules use the same amino acid vocabulary, we share their initial linear projection to establish a common residue basis. Since physicochemical descriptors are an established way to represent protein sequences~\citep{yang2018learned}, we incorporate four binary indicators to make selected residue properties explicit alongside the core amino acid identity. Specifically, each residue is represented by a 25-dimensional feature vector: a 21-dimensional one-hot encoding (for the 20 standard amino acids and the unknown symbol $\mathtt{X}$), concatenated with four binary physicochemical indicators (aromatic, acidic, basic, and hydrophilic; \Cref{app:protein-seq}). Characters outside the standard alphabet are mapped to $\mathtt{X}$ (assigned zero-valued physicochemical indicators), while the aforementioned padding positions use strictly all-zero feature vectors. These combined features form $\mathbf{F}$ in \Cref{eq:residue-representation}. The same bias-free projection $\mathbf{W}_{\mathrm{in}}$ is then applied across all three inputs, mapping from $d_{\mathrm{in}}=25$ to $d_{\mathrm{model}}=64$ (\Cref{fig:Model_Structure}(b)).

\noindent\textbf{Molecule-specific contextual encoding.} Building on this shared residue basis, molecule-specific encoders then capture the distinct sequence contexts of the peptide, MHC, and TCR. Sinusoidal positional encodings $\mathbf{P}$ (\Cref{app:pe}) are added after the shared projection to form $\mathbf{H}^{(0)}$. The three sequences then pass through separate self-attention and feed-forward blocks with independent parameters. These attention layers use the multi-head form of the $\operatorname{SelfAttn}$ operation defined in \Cref{eq:self-attn}, with $h=4$ and $d=16$. The resulting encoder outputs are denoted by $\mathbf{H}_p$, $\mathbf{H}_M$, and $\mathbf{H}_T$.

\subsection{Two-Stage Fusion and Transfer}
\label{sec:dissection}

Following the paradigm of pMTnet~\citep{lu2021pMTnet}, Fusion-pMT uses peptide--MHC pretraining to transfer pairwise binding knowledge into the tripartite prediction task. However, whereas prior methods typically compress this knowledge into a single-vector bottleneck (discarding positional detail for receptor-attention queries), Fusion-pMT explicitly disentangles the prediction and interaction roles of the pMHC representation (\Cref{fig:Model_Structure}(a)). A global summary supports pairwise prediction, while the retained continuous tokens carry the residue-level MHC context into position-specific interaction with the TCR sequence.

\textbf{Stage 1: Pretraining via peptide--MHC binding.}
We first pretrain the peptide and MHC modules on established peptide--MHC binding datasets~\citep{chu2022transphla}. To condition the peptide representation on its presenting MHC context, we apply the multi-head form of $\operatorname{CrossAttn}(\mathbf{H}_p,\mathbf{H}_M)$ (\Cref{eq:cross-attn}), using peptide representations as queries and MHC representations as keys and values (\Cref{fig:Model_Structure}(c)). A self-attention block refines the resulting sequence into contextualized pMHC tokens $\mathbf{H}_{\mathrm{pM}}$, which are subsequently aggregated into a global pMHC summary $\mathbf{s}_{\mathrm{pM}}$.

\textbf{Stage 2: Fine-tuning for peptide--MHC--TCR interaction prediction.}
To effectively transfer the learned pairwise representations and stabilize early training, we employ a two-phase fine-tuning strategy. We initialize the peptide--MHC modules with the Stage 1 weights and temporarily freeze them, allowing the newly initialized TCR and fusion modules to stabilize. We then unfreeze the entire network for end-to-end joint adaptation using differential learning rates.

With this transfer in place, Stage~2 uses the retained pMHC token sequence for receptor-specific fusion (\Cref{fig:Model_Structure}(a)). The block applies the multi-head form of $\operatorname{CrossAttn}(\mathbf{H}_{\mathrm{pM}},\mathbf{H}_T)$, mapping each MHC-conditioned peptide position as a query over TCR keys and values. This directed attention formulation preserves the peptide sequence dimension across both fusion stages while reflecting the biological prior that MHC presentation precedes receptor interaction. A self-attention block refines the resulting sequence into $\mathbf{H}_{\mathrm{pMT}}$, yielding the joint summary 
{$\mathbf{s}_{\mathrm{joint}}$} via a second readout.

\subsection{Class-Attention Readout and Prediction}
\label{sec:readout-prediction}

{To extract these aforementioned summaries ($\mathbf{s}_{\mathrm{pM}}$ and $\mathbf{s}_{\mathrm{joint}}$) from their respective sequence states ($\mathbf{H}_{\mathrm{pM}}$ and $\mathbf{H}_{\mathrm{pMT}}$), we apply the multi-head form of class attention introduced in \Cref{sec:ca-in-bio}. Two independently parameterized readout modules aggregate the spatial context using dedicated learned CLS tokens.}

\textbf{Dual-summary prediction.}
Because the pMHC summary captures the peptide--MHC state prior to TCR interaction, concatenating it with the joint summary provides the final classifier with explicit access to both hierarchical stages of representation. Letting $g_{\mathrm{pM}}$ and $g_{\mathrm{pMT}}$ denote the linear classification heads and $\sigma$ the sigmoid function, the prediction scores are formulated as $\hat{y}_{\mathrm{pM}}=\sigma(g_{\mathrm{pM}}(\mathbf{s}_{\mathrm{pM}}))$ and $\hat{y}_{\mathrm{pMT}}=\sigma(g_{\mathrm{pMT}}([\mathbf{s}_{\mathrm{pM}};\mathbf{s}_{\mathrm{joint}}]))$, where the brackets denote feature concatenation. (Complete architectural details, including normalizations and intermediate dimensions, are provided in \Cref{sec:imple-details}).

\section{Experimental Results and Analysis}
\label{sec:exp}

In this section, we first outline the 
{experimental setup and evaluation protocol} in \Cref{exp:set}, followed by the 
{baseline models} in \Cref{exp:bas}. We then evaluate our framework across two core tasks: peptide--MHC binding (\Cref{exp:imm}) and peptide--MHC--TCR interaction prediction (\Cref{exp:pmt}). Finally, \Cref{exp:abl} presents 
{architectural and pretraining ablations alongside a progressive comparison with pMTnet}.

\subsection{
{Experimental Setup and Evaluation}} \label{exp:set}

We evaluate the two stages of our framework on their respective benchmark tasks: \textbf{Fusion-pM} for peptide--MHC binding and \textbf{Fusion-pMT} for peptide--MHC--TCR interaction prediction. Both models were trained using the AdamW optimizer with a batch size of 1,024 across five random seeds. Stage~1 was trained with a learning rate of 0.002 for up to 80 epochs, whereas Stage~2 used a learning rate of 0.0015 for up to 100 epochs. Complete architectural and optimization hyperparameters are summarized in \Cref{tab:training_current}.

\begin{table}[!ht]
\centering
\caption{
{Architecture and training hyperparameters for Stage~1 (Fusion-pM) and Stage~2 (Fusion-pMT) across five random seeds (4001, 4003, 4007, 4013, and 4019). Parameter counts refer to the instantiated models; the Stage~2 count includes its transferred modules.}}
\label{tab:training_current}
{
\begin{tabular}{lcc}
\toprule
\textbf{Setting} & \textbf{Stage 1} & \textbf{Stage 2} \\
\midrule
Representation dimension & 64 & 64 \\
Attention heads & 4 & 4 \\
FFN hidden dimension & 256 & 256 \\
Dropout & 0.1 & 0.1 \\
Model parameters & 252,033 & 452,546 \\
Optimizer & AdamW & AdamW \\
Initial learning rate & 0.002 & 0.0015 \\
Weight decay & 0.0001 & 0.001 \\
Training batch size & 1,024 & 1,024 \\
Evaluation batch size & 4,096 & 4,096 \\
Maximum epochs & 80 & 100 \\
Minimum epochs & 30 & 30 \\
Early-stopping patience & 15 & 15 \\
Warmup epochs & 1 & 1 \\
Transferred-module freeze epochs & -- & 10 \\
Transferred-module learning-rate scale & -- & 0.3 \\
Loss & BCEWithLogits & BCEWithLogits \\
\bottomrule
\end{tabular}}
\end{table}

\noindent\textbf{Peptide--MHC binding.} For Stage~1 training and evaluation, we adopt the dataset partitions from TransPHLA~\citep{chu2022transphla}, using the pre-compiled MHC pseudosequences provided therein. We refer to TransPHLA's \texttt{Independent} testing partition as the \textbf{pMHC testing dataset} throughout this paper. Because this benchmark is naturally balanced, comprising 171,438 examples (85,876 positives and 85,562 negatives), all pMHC evaluation metrics are computed without class reweighting.

\noindent\textbf{Peptide--MHC--TCR interaction prediction.} For Stage~2 training and validation, we follow the data setup established by pMTnet and subsequent studies~\citep{lu2021pMTnet, han2023pan}, where each known positive interaction is paired with ten synthetically generated negatives and oversampled tenfold during training to balance the classes. We refer to the fixed pMTnet-derived test set of 6,809 examples (619 positives and 6,190 negatives) as the \textbf{pMT testing dataset}. The positive examples span 272 distinct TCR CDR3 sequences, 224 peptides, and 24 MHC types. For class-balanced evaluation on this 1:10 test set, positive and negative test examples are weighted 10 and 1, respectively, across all compared models to give both classes equal total contribution. {To resolve MHC identifiers into pseudosequences, we query the reference mapping sequentially by exact allele, two-field match, and the first lexicographically ordered allele within the same group; unresolved identifiers map to 34 $\mathtt{X}$ symbols.}

{\textbf{Evaluation protocol and checkpoint selection.} We evaluate model performance using five standard metrics: Accuracy (ACC), F1 Score, Matthews Correlation Coefficient (MCC), Receiver Operating Characteristic Area Under the Curve (ROC AUC), and Precision--Recall Area Under the Curve (PR AUC). Across both tasks, we compute PR AUC as average precision; the figures show the mean and standard error of the mean (SEM) across training runs, while the tables report the sample standard deviation (SD). For each run, we select the Stage~1 checkpoint that minimizes validation binary cross-entropy and the Stage~2 checkpoint that maximizes validation average precision. We evaluate threshold-dependent test metrics using a fixed decision threshold chosen to maximize unweighted MCC on the validation set, while test PR AUC is computed directly from the continuous prediction scores under the task-specific evaluation weights. Statistical comparison protocols across seeded runs and fixed checkpoints are detailed in \Cref{app:seed-comparisons}.}

\subsection{
    {Baseline Models}
}
\label{exp:bas}

To establish a rigorous evaluation landscape, we benchmark our proposed framework against a diverse set of established models across both hierarchical stages of interaction. For the Stage~1 peptide--MHC (pMHC) binding task, we compare against several strong baselines, including TransPHLA~\citep{chu2022transphla}, NetMHCpan, ANN~{\citep{andreatta2016gapped}}, and PickPocket~\citep{zhang2009pickpocket}. To ensure robust comparisons, we report both a retrained TransPHLA model and the authors' released checkpoint, whereas ANN and PickPocket are evaluated exclusively using their official predictors provided through their respective official prediction services, as their native training pipelines are not publicly accessible. For the NetMHCpan baseline, we faithfully adopt the exact implementation integrated within the pMTnet framework~\citep{lu2021pMTnet}, coupling its frozen LSTM-based pMHC encoder with a retrained linear classifier.

{For the Stage~2 peptide--MHC--TCR interaction prediction task, we benchmark Fusion-pMT against four baseline methods:} pMTnet~\citep{lu2021pMTnet}, PISTE~\citep{feng2024sliding}, an ESM-2-based MLP~\citep{lin2023evolutionary}, and ERGO II~\citep{springer2021ergo2}. These baselines reflect diverse modeling approaches, including LSTM-based architectures, sliding and cross-attention mechanisms, and large-scale pretrained protein language models. For the established task-specific architectures (pMTnet, PISTE, and ERGO II), we separately report performance for both the authors' released checkpoints (yielding fixed predictions) and models retrained in-house (averaged across five seeds). 
Conversely, the ESM-2 baseline serves as a dedicated representation-learning probe, which concatenates fine-tuned ESM-2 8M representations into an MLP classifier. 
Both PISTE variants were evaluated on 5,656 examples and the released ERGO II checkpoint on 6,631 examples. All other pMT configurations were evaluated on the full test set of 6,809 examples.

\begin{figure}[!tb]
    \centering
    \includegraphics[width=0.99\textwidth]{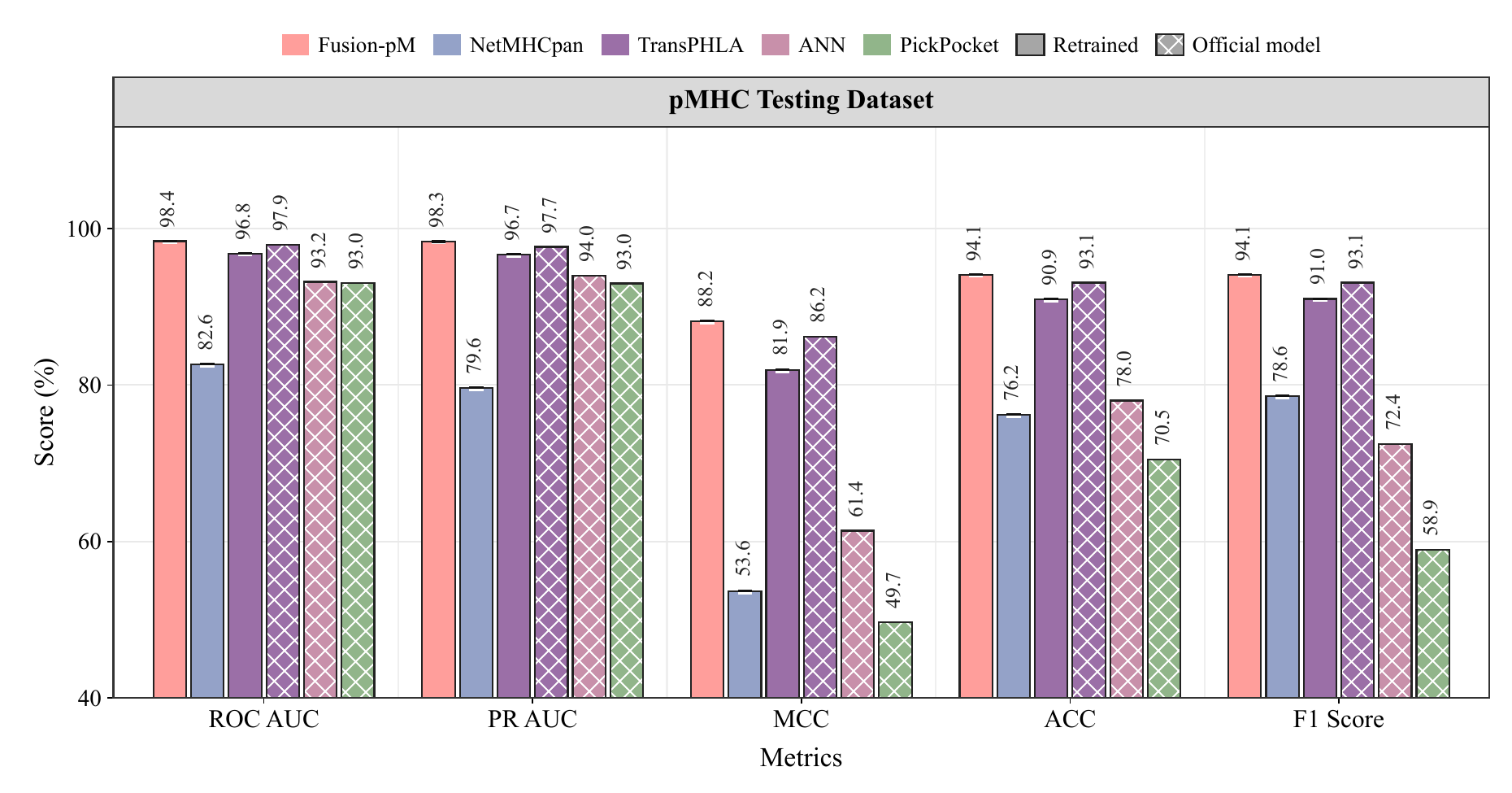}
    \caption{\textbf{Peptide--MHC binding performance.} Plain bars represent retrained models, and crosshatched bars indicate released official checkpoints. All scores use the native class proportions without reweighting, with error bars showing SEM over five runs. Notably, the NetMHCpan baseline evaluates a retrained linear classifier atop the frozen pMTnet pMHC encoder. Model definitions and test coverage are detailed in \Cref{exp:bas}, with complete values in Appendix~\ref{app:baseline-results}.
}
    \label{fig:ModelPerformance}
\end{figure}

\subsection{Peptide--MHC Binding Prediction}
\label{exp:imm}

We first evaluate Fusion-pM as a standalone peptide--MHC predictor. Establishing strong performance at this initial stage is crucial for validating our core architectural choices, such as the shared residue projection and cross-attention fusion.

On the pMHC testing dataset, Fusion-pM achieves a mean PR AUC of 98.33\% and ROC AUC of 98.40\% (\Cref{fig:ModelPerformance,tab:baseline_current}). The corresponding scores are 96.69\% and 96.82\% for retrained TransPHLA, and 97.65\% and 97.89\% for its released checkpoint. Fusion-pM also attains higher mean MCC (88.17\%), accuracy (94.08\%), and F1 score (94.10\%) than both TransPHLA versions. By comparison, the NetMHCpan baseline yields a mean PR AUC of 79.58\% and ROC AUC of 82.64\%. Ultimately, these empirical results do more than establish Fusion-pM as an accurate pairwise predictor; they confirm that our first-stage architecture successfully extracts the high-quality binding representations required to anchor the downstream peptide--MHC--TCR interaction modeling.

\subsection{Peptide--MHC--TCR Interaction Prediction}
\label{exp:pmt}

\begin{figure*}[t!]
    \centering
    \includegraphics[width=0.99\textwidth]{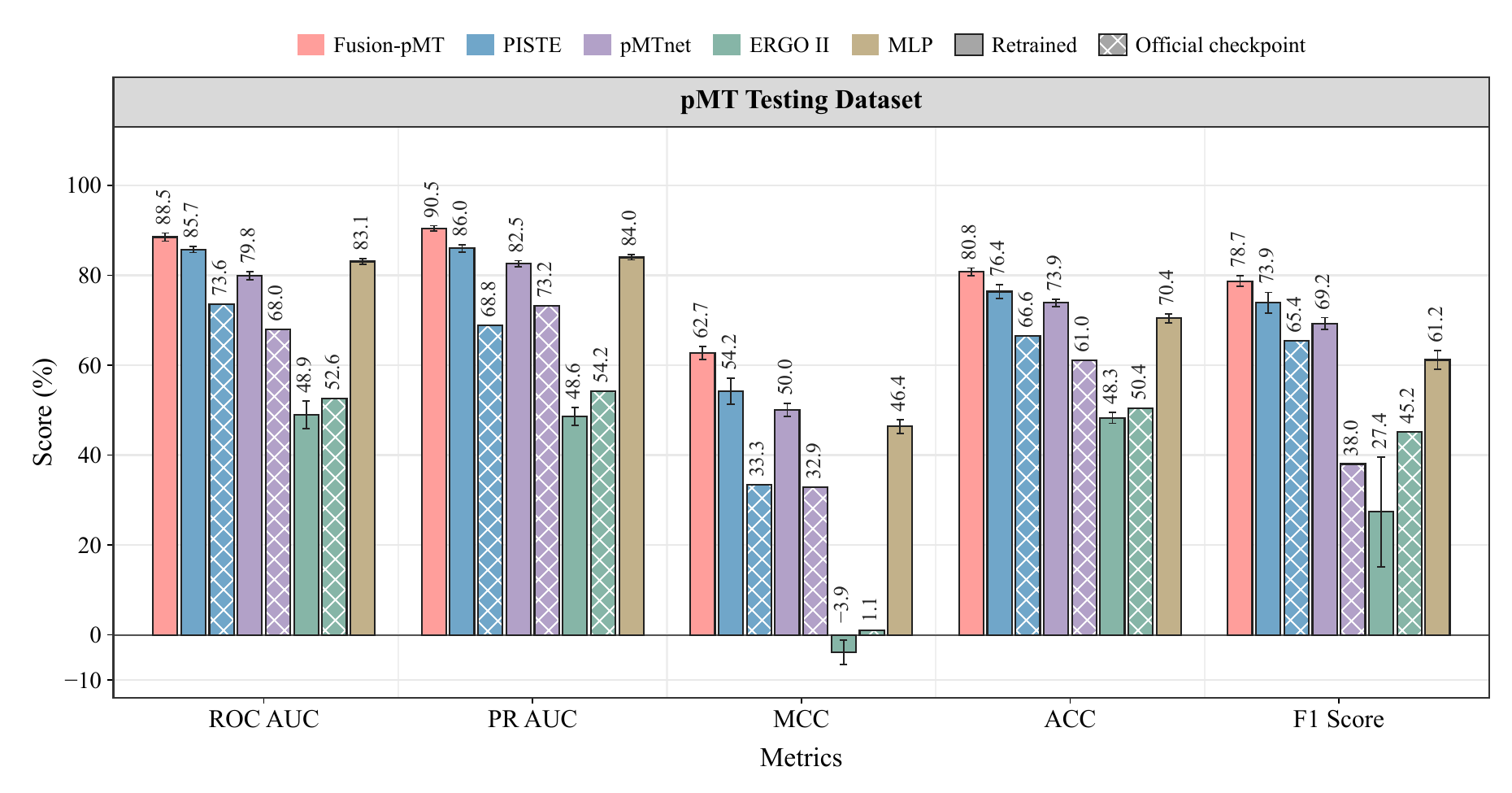}
    \caption{\textbf{Tripartite interaction prediction.} 
    Plain bars represent retrained models, and crosshatched bars indicate released official model checkpoints. Error bars show SEM over five runs. Model definitions and test coverage are detailed in \Cref{exp:bas}.}
    \label{fig:Model Performance}
\end{figure*}

\begin{table}[!tbp]
\centering
\footnotesize
\caption{\textbf{Baseline performance on the pMT testing dataset (\%).} Retrained scores represent the mean $\pm$ sample SD over five independent runs; released checkpoints yield a single deterministic score. All scores apply the class-balanced evaluation weights defined in \Cref{exp:set}. Comprehensive results for both tasks are detailed in Appendix~\ref{app:baseline-results}.
}
\label{tab:pmt_baselines_main}
\setlength{\tabcolsep}{3pt}
\resizebox{\textwidth}{!}{\begin{tabular}{llccccc}
\toprule
\textbf{Model} & \textbf{Version} & \textbf{ROC AUC} & \textbf{PR AUC} & \textbf{MCC} & \textbf{ACC} & \textbf{F1} \\
\midrule
\textbf{Fusion-pMT} & Retrained & $88.48\pm2.03$ & $90.46\pm1.30$ & $62.69\pm3.31$ & $80.76\pm1.92$ & $78.70\pm2.70$ \\
PISTE & Retrained & $85.72\pm1.48$ & $86.00\pm1.78$ & $54.19\pm6.38$ & $76.37\pm3.58$ & $73.91\pm5.13$ \\
PISTE & Checkpoint & $73.61$ & $68.85$ & $33.34$ & $66.55$ & $65.40$ \\
pMTnet & Retrained & $79.85\pm2.02$ & $82.54\pm1.53$ & $49.98\pm3.24$ & $73.86\pm1.88$ & $69.25\pm2.99$ \\
pMTnet & Checkpoint & $68.02$ & $73.15$ & $32.88$ & $61.02$ & $38.02$ \\
ERGO II & Retrained & $48.93\pm6.78$ & $48.62\pm4.41$ & $-3.91\pm6.13$ & $48.27\pm2.74$ & $27.36\pm27.33$ \\
ERGO II & Checkpoint & $52.55$ & $54.20$ & $1.08$ & $50.39$ & $45.15$ \\
MLP & Retrained & $83.05\pm1.51$ & $83.97\pm1.28$ & $46.36\pm3.43$ & $70.43\pm2.29$ & $61.16\pm4.53$ \\
\bottomrule
\end{tabular}}
\end{table}

We next evaluate whether the complete two-stage framework improves tripartite interaction prediction relative to established predictors and a pretrained protein language model baseline (\Cref{fig:Model Performance,tab:pmt_baselines_main}). Across all 6,809 testing examples, Fusion-pMT achieves a mean PR AUC of 90.46\% and ROC AUC of 88.48\%, exceeding the controlled \texttt{pMTnet-retrained} comparator (82.54\% PR AUC, 79.85\% ROC AUC) by 7.92 percentage points in PR AUC, and the fine-tuned ESM-2 8M \texttt{MLP} baseline (83.97\% PR AUC, 83.05\% ROC AUC) by 6.49 points. Retrained ERGO II reaches 48.62\% PR AUC and 48.93\% ROC AUC under this protocol, whereas the official pMTnet checkpoint yields a fixed 73.15\% PR AUC and 68.02\% ROC AUC. These empirical gains validate our full framework, demonstrating the value of carrying pretrained pMHC representations into residue-level TCR interaction modeling.

Across the four baseline configurations evaluated on the full test set,
all 20 metric comparisons yielded unadjusted $p<0.01$. For PR AUC, the
paired $t$-test against retrained pMTnet gave $p=4.0\times10^{-4}$,
and Welch's $t$-test against the fine-tuned ESM-2 8M MLP gave
$p=2.3\times10^{-5}$ (Table~\ref{tab:model_comparison},
Appendix~\ref{app:seed-comparisons}). Furthermore, at the validation-selected decision threshold, Fusion-pMT achieves 71.31\% sensitivity, 90.21\% specificity, and 87.93\% weighted precision. 
{The corresponding pMHC baseline results are reported in Appendix~\ref{app:baseline-results}.}

\subsection{Ablation Studies and Progressive Comparison}
\label{exp:abl}
\label{sec: same embeddings}

\begin{figure}[!tb]
    \centering
    \begin{subfigure}[t]{0.49\linewidth}
        \centering
        \includegraphics[width=\linewidth]{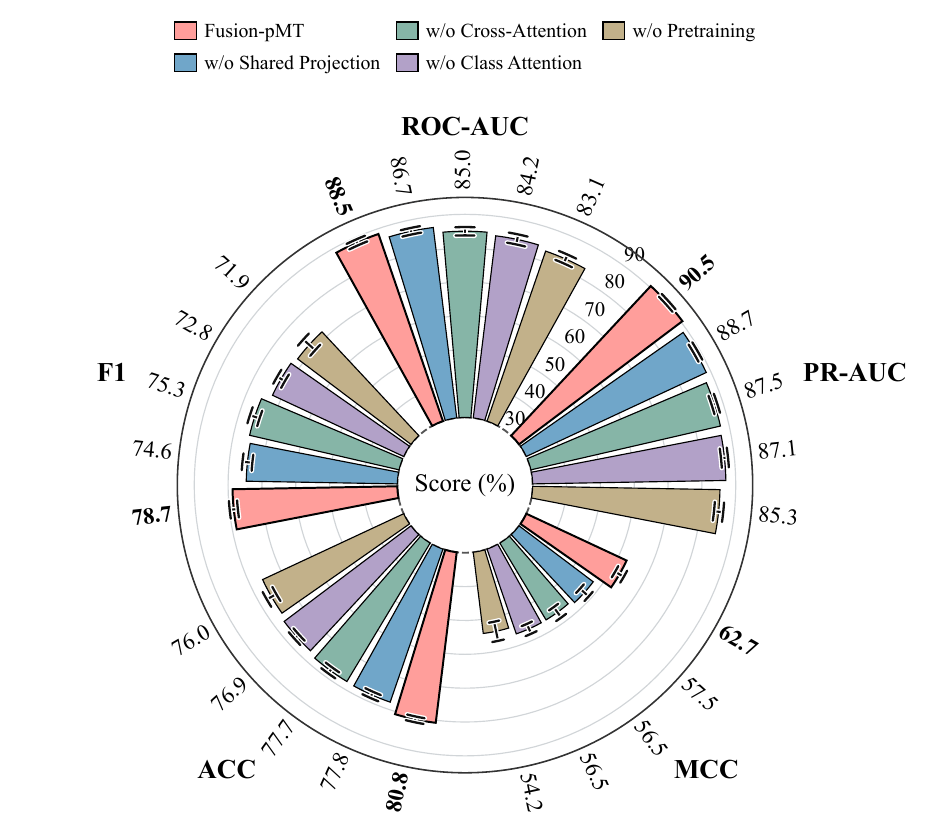}
        \caption{}
    \end{subfigure}\hfill
    \begin{subfigure}[t]{0.49\linewidth}
        \centering
        \includegraphics[width=\linewidth]{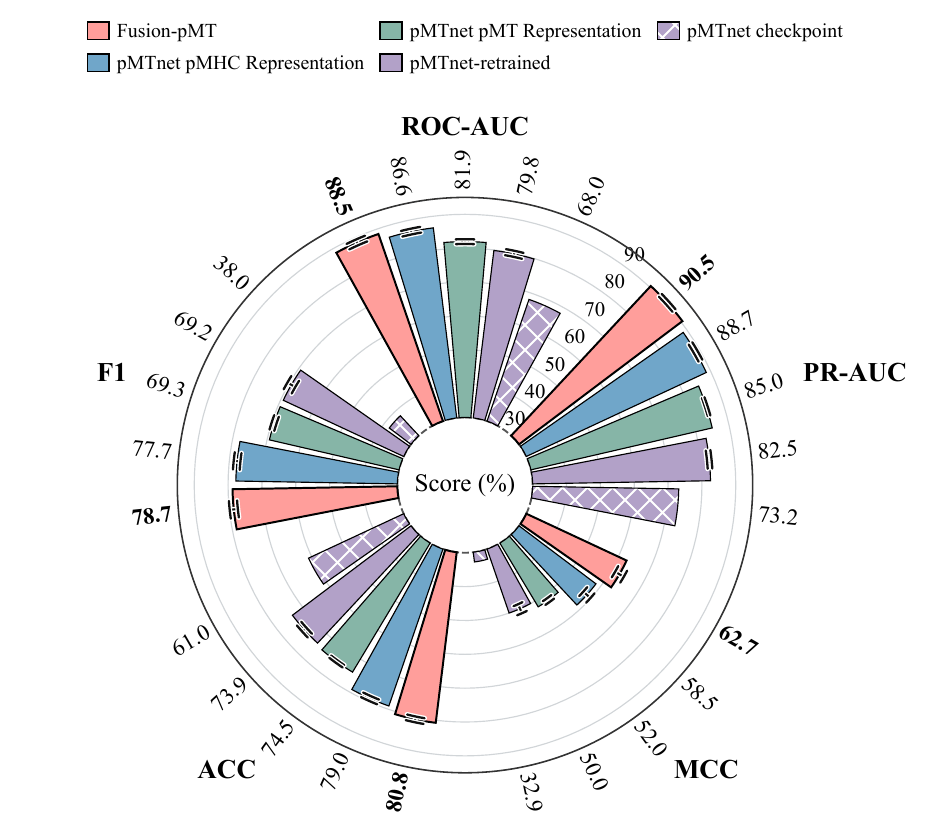}
        \caption{}
    \end{subfigure}
    \caption{
    \textbf{
    {Ablation studies of Fusion-pMT and progressive comparison with pMTnet.}} 
    (a)~Performance impact of removing core mechanisms: the shared input projection, cross-attention fusion, Stage~2 class-attention readout, and Stage~1 pretraining. 
    (b) 
    {Progressive architectural replacements mapping the performance trajectory from Fusion-pMT to pMTnet.} Bars represent means with SEM over five independent runs, excluding the fixed deterministic pMTnet checkpoint. 
    {Detailed configuration settings are provided in \Cref{exp:abl}.}}
    \label{fig:Model Performance2}
\end{figure}

To isolate the drivers of Fusion-pMT's performance, in \Cref{fig:Model Performance2,tab:ablation_current}, we conduct two complementary analyses. First, component ablations quantify the impact of our core architectural and transfer-learning choices. Second, a progressive structural comparison traces how intermediate design shifts bridge the gap toward the representative pMTnet baseline.

\noindent\textbf{Architectural and pretraining ablations.}
Fusion-pMT relies on a combination of shared residue representations, hierarchical attention mechanisms, and staged transfer learning to capture complex molecular interactions. To isolate the independent contribution of each core design choice, we evaluate four targeted ablations against the full model's 90.46\% mean PR AUC baseline (\Cref{fig:Model Performance2}(a)). 
\ding{182}~First, \texttt{w/o Shared Projection} uses independent sequence projections, reducing PR AUC by 1.80 percentage points {and confirming the benefit of a shared residue basis.} 
\ding{183}~Second, \texttt{w/o Cross-Attention} substitutes cross-attention blocks with approximately parameter-matched MLPs; this drops PR AUC by 3.00 points, {demonstrating the superiority of attention-based residue interactions.}
\ding{184}~Third, \texttt{w/o Class Attention} replaces the Stage~2 readout with a capacity-matched MLP and mean pooling (\Cref{app:ablation-results}), lowering PR AUC by 3.37 points and validating class-attention aggregation. 
\ding{185}~Lastly, \texttt{w/o Pretraining} trains the network from scratch without Stage~1 transfer. This causes the largest drop (5.12 points to 85.34\%), {highlighting the critical role of pairwise binding priors.} 
All four ablations yielded lower mean PR AUC than the full model, with
unadjusted one-sided paired $t$-test $p$-values between 0.0182 and 0.0439
(Table~\ref{tab:model_comparison}, Appendix~\ref{app:seed-comparisons}).

\noindent\textbf{Progressive comparison with pMTnet.}
Both Fusion-pMT and pMTnet share a foundational two-stage transfer learning paradigm, leveraging peptide--MHC pretraining to inform downstream tripartite predictions. Despite this shared conceptual framework, a substantial performance gap exists between them. To systematically trace how specific technical design choices drive this divergence, we cumulatively replace our components with pMTnet configurations (\Cref{fig:Model Performance2}(b), \Cref{tab:ablation_current}). 
\ding{182}~First, \texttt{pMTnet pMHC Representation} substitutes Stage~1 features with frozen pMTnet pMHC embeddings and an adapter, reducing mean PR AUC to 88.68\%. 
\ding{183}~Second, \texttt{pMTnet pMT Representation} additionally replaces the TCR encoder and classifier with pMTnet's frozen features, concatenation, and MLP. 
{This triggers the largest stepwise decrease (3.71 points), dropping PR AUC to 84.97\%.}
\ding{184}~Third, \texttt{pMTnet-retrained} maintains this architecture but adopts pMTnet's pairwise ranking loss and regularization, yielding a mean PR AUC of 82.54\%.
Together, these successive replacements trace the 7.92-percentage-point gap in mean PR AUC between Fusion-pMT and \texttt{pMTnet-retrained}.
The released \texttt{pMTnet checkpoint} provides a fixed external reference.

\begin{table*}[tbp]
\centering
\caption{
{\textbf{Comprehensive ablation and progressive replacement results (\%).} Retrained scores represent the mean $\pm$ sample SD over five independent runs, whereas the released pMTnet checkpoint yields a single deterministic score. 
}}
\label{tab:ablation_current}
\small
\begin{tabular*}{\textwidth}{@{\extracolsep{\fill}}lccccc}
\toprule
\textbf{Condition} & \textbf{ROC AUC} & \textbf{PR AUC} & \textbf{MCC} & \textbf{ACC} & \textbf{F1} \\
\midrule
\multicolumn{6}{c}{\textbf{\Cref{fig:Model Performance2}(a): architecture and training ablations}} \\
\addlinespace[2pt]
\textbf{Fusion-pMT} & $88.48\pm2.03$ & $90.46\pm1.30$ & $62.69\pm3.31$ & $80.76\pm1.92$ & $78.70\pm2.70$ \\
w/o Shared Projection & $86.68\pm2.17$ & $88.66\pm1.43$ & $57.45\pm3.83$ & $77.83\pm2.22$ & $74.64\pm3.27$ \\
w/o Cross-Attention & $85.01\pm2.75$ & $87.46\pm1.89$ & $56.47\pm4.24$ & $77.71\pm2.36$ & $75.32\pm3.37$ \\
w/o Class Attention & $84.20\pm3.37$ & $87.09\pm2.34$ & $56.49\pm3.54$ & $76.92\pm2.01$ & $72.76\pm2.94$ \\
w/o Pretraining & $83.12\pm3.17$ & $85.34\pm3.06$ & $54.16\pm6.46$ & $75.99\pm3.60$ & $71.93\pm5.27$ \\
\midrule
\multicolumn{6}{c}{\textbf{\Cref{fig:Model Performance2}(b): progressive replacement}} \\
\addlinespace[2pt]
\textbf{Fusion-pMT} & $88.48\pm2.03$ & $90.46\pm1.30$ & $62.69\pm3.31$ & $80.76\pm1.92$ & $78.70\pm2.70$ \\
pMTnet pMHC Representation & $86.57\pm1.91$ & $88.68\pm1.41$ & $58.46\pm3.17$ & $79.02\pm1.68$ & $77.71\pm2.23$ \\
pMTnet pMT Representation & $81.94\pm1.58$ & $84.97\pm0.88$ & $51.96\pm1.48$ & $74.47\pm0.95$ & $69.31\pm1.63$ \\
pMTnet-retrained & $79.85\pm2.02$ & $82.54\pm1.53$ & $49.98\pm3.24$ & $73.86\pm1.88$ & $69.25\pm2.99$ \\
pMTnet checkpoint & $68.02$ & $73.15$ & $32.88$ & $61.02$ & $38.02$ \\
\bottomrule
\end{tabular*}
\end{table*}

\paragraph{Two-stage fusion with pretrained protein representations.}
We evaluated Fusion-pMT with frozen and fine-tuned ESM-2 8M and 35M
residue representations (Appendix~\ref{app:esm2_comparison},
Table~\ref{tab:esm2_comparison}). Across all four settings, the
Fusion-pMT pipeline achieved higher mean scores on all five metrics than
the corresponding pooled concatenation baseline (\texttt{Concat} + MLP).
The largest mean PR AUC gains were obtained with frozen backbones:
16.71 percentage points for ESM-2 8M (86.86\% vs.\ 70.15\%) and
12.05 points for ESM-2 35M (87.87\% vs.\ 75.82\%). Notably, none of the
ESM-2 variants surpassed the original Fusion-pMT (90.46\% PR AUC under
its standard training schedule), showing that strong performance does not
require heavy pretrained sequence backbones.

\section{Conclusion and Discussion}
\label{sec:conclusion}

We presented Fusion-pMT, a biologically motivated framework for predicting peptide--MHC--TCR interactions. Our core insight is that preserving the sequence formulation of MHC-conditioned peptides, rather than compressing them into single-vector bottlenecks, is critical for effective transfer to residue-level TCR interaction modeling. By explicitly aligning our architecture with the sequential stages of immune recognition, we naturally obtain a highly competitive standalone peptide--MHC predictor in Stage~1.

Empirically, Fusion-pMT significantly outperforms established baselines across the full pMT test set (e.g., achieving a 7.92 percentage point gain in mean class-balanced PR AUC over \texttt{pMTnet-retrained}). 

{Furthermore, supplementary ESM-2 comparisons demonstrate that our residue-level two-stage fusion and pretraining strategies adapt effectively to large pretrained protein representations.} Ultimately, this work provides a scalable representation-learning foundation for capturing complex, higher-order molecular interactions; detailed discussions on model limitations and broader impacts are provided in \Cref{app:limitations}.

\newpage
\bibliography{main-arxiv}

\begin{thebibliography}{48}
\providecommand{\natexlab}[1]{#1}
\providecommand{\url}[1]{\texttt{#1}}
\expandafter\ifx\csname urlstyle\endcsname\relax
  \providecommand{\doi}[1]{doi: #1}\else
  \providecommand{\doi}{doi: \begingroup \urlstyle{rm}\Url}\fi

\bibitem[Andreatta \& Nielsen(2016)Andreatta and Nielsen]{andreatta2016gapped}
Massimo Andreatta and Morten Nielsen.
\newblock {Gapped sequence alignment using artificial neural networks: application to the MHC class I system}.
\newblock \emph{Bioinformatics}, 32\penalty0 (4):\penalty0 511--517, 2016.
\newblock ISSN 1367-4811.
\newblock \doi{10.1093/bioinformatics/btv639}.
\newblock URL \url{https://doi.org/10.1093/bioinformatics/btv639}.

\bibitem[Arieta et~al.(2023)Arieta, Xie, Rothenberg, Diao, Harjanto, Meda, Marquart, Koenitzer, Sciuto, Lobo, Zuiani, Krumm, Cadima~Couto, Hein, Heinen, Ziegenhals, Liu-Lupo, Vogel, Srouji, Fesser, Thanki, Walzer, Addona, T{\"u}reci, {\c{S}}ahin, Gaynor, and Poran]{Arieta2023BNT162b4}
Christina~M. Arieta, Yushu~Joy Xie, Daniel~A. Rothenberg, Huitian Diao, Dewi Harjanto, Shirisha Meda, Krisann Marquart, Byron Koenitzer, Tracey~E. Sciuto, Alexander Lobo, Adam Zuiani, Stefanie~A. Krumm, Carla~Iris Cadima~Couto, Stephanie Hein, Andr{\'e}~P. Heinen, Thomas Ziegenhals, Yunpeng Liu-Lupo, Annette~B. Vogel, John~R. Srouji, Stephanie Fesser, Kaushik Thanki, Kerstin Walzer, Theresa~A. Addona, {\"O}zlem T{\"u}reci, U{\u{g}}ur {\c{S}}ahin, Richard~B. Gaynor, and Asaf Poran.
\newblock {The T-cell-directed vaccine BNT162b4 encoding conserved non-spike antigens protects animals from severe SARS-CoV-2 infection}.
\newblock \emph{Cell}, 186\penalty0 (11):\penalty0 2392--2409.e21, 2023.
\newblock ISSN 0092-8674.
\newblock \doi{10.1016/j.cell.2023.04.007}.
\newblock URL \url{https://doi.org/10.1016/j.cell.2023.04.007}.

\bibitem[Bradley(2023)]{bradley2023structure}
Philip Bradley.
\newblock Structure-based prediction of t cell receptor: peptide-mhc interactions.
\newblock \emph{Elife}, 12:\penalty0 e82813, 2023.

\bibitem[Chen et~al.(2021)Chen, Fan, and Panda]{chen2021crossvit}
Chun-Fu~Richard Chen, Quanfu Fan, and Rameswar Panda.
\newblock Crossvit: Cross-attention multi-scale vision transformer for image classification.
\newblock In \emph{{Proceedings of the IEEE/CVF International Conference on Computer Vision}}, pp.\  357--366, 2021.

\bibitem[Chen et~al.(2022)Chen, Hazarika, Namazifar, Liu, Jin, and Hakkani-Tur]{chen-etal-2022-empowering}
Yifan Chen, Devamanyu Hazarika, Mahdi Namazifar, Yang Liu, Di~Jin, and Dilek Hakkani-Tur.
\newblock Empowering parameter-efficient transfer learning by recognizing the kernel structure in self-attention.
\newblock In Marine Carpuat, Marie-Catherine de~Marneffe, and Ivan~Vladimir Meza~Ruiz (eds.), \emph{Findings of the Association for Computational Linguistics: NAACL 2022}, pp.\  1375--1388, Seattle, United States, July 2022. Association for Computational Linguistics.
\newblock \doi{10.18653/v1/2022.findings-naacl.102}.
\newblock URL \url{https://aclanthology.org/2022.findings-naacl.102}.

\bibitem[Chi et~al.(2024)Chi, Pepper, and Thomas]{Chi2024cell}
Hongbo Chi, Marion Pepper, and Paul~G. Thomas.
\newblock Principles and therapeutic applications of adaptive immunity.
\newblock \emph{Cell}, 187\penalty0 (9):\penalty0 2052--2078, 2024.
\newblock ISSN 0092-8674.
\newblock \doi{10.1016/j.cell.2024.03.037}.
\newblock URL \url{https://doi.org/10.1016/j.cell.2024.03.037}.

\bibitem[Chu et~al.(2021)Chu, Zhang, Wang, Zhang, Wang, Wang, Wang, Jiang, Salahub, Xiong, and Wei]{chu2021transmut}
Yanyi Chu, Yan Zhang, Qiankun Wang, Lingfeng Zhang, Xuhong Wang, Yanjing Wang, Jianmin Wang, Xue Jiang, Dennis Salahub, Yi~Xiong, and Dong-qing Wei.
\newblock {TransMut: a program to predict HLA-I peptide binding and optimize mutated peptides for vaccine design by the Transformer-derived self-attention model}.
\newblock \emph{{Research Square}}, 2021.
\newblock \doi{10.21203/rs.3.rs-785618/v1}.
\newblock URL \url{https://doi.org/10.21203/rs.3.rs-785618/v1}.

\bibitem[Chu et~al.(2022)Chu, Zhang, Wang, Zhang, Wang, Wang, Salahub, Xu, Wang, Jiang, Xiong, and Wei]{chu2022transphla}
Yanyi Chu, Yan Zhang, Qiankun Wang, Lingfeng Zhang, Xuhong Wang, Yanjing Wang, Dennis~Russell Salahub, Qin Xu, Jianmin Wang, Xue Jiang, Yi~Xiong, and Dong-qing Wei.
\newblock {A transformer-based model to predict peptide--HLA class I binding and optimize mutated peptides for vaccine design}.
\newblock \emph{Nature Machine Intelligence}, 4\penalty0 (3):\penalty0 300--311, 2022.
\newblock ISSN 2522-5839.
\newblock \doi{10.1038/s42256-022-00459-7}.
\newblock URL \url{https://doi.org/10.1038/s42256-022-00459-7}.

\bibitem[Dong et~al.(2022)Dong, Pi, Bartels-Burgahn, Saltukoglu, Liang, Yang, Alt, Reth, and Wu]{dong2022bcrassembly}
Ying Dong, Xiong Pi, Frauke Bartels-Burgahn, Deniz Saltukoglu, Zhuoyi Liang, Jianying Yang, Frederick~W. Alt, Michael Reth, and Hao Wu.
\newblock Structural principles of b cell antigen receptor assembly.
\newblock \emph{Nature}, 612\penalty0 (7938):\penalty0 156--161, 2022.
\newblock \doi{10.1038/s41586-022-05412-7}.

\bibitem[D’Angelo et~al.(2024)D’Angelo, Araujo, Razak, Agulnik, Attia, Blay, Garcia, Charlson, Choy, Demetri, Druta, Forcade, Ganjoo, Glod, Keedy, Cesne, Liebner, Moreno, Pollack, Schuetze, Schwartz, Strauss, Tap, Thistlethwaite, Morales, Wagner, Wilky, McAlpine, Hudson, Navenot, Wang, Bai, Rafail, Wang, Sun, Fernandes, Van~Winkle, Elefant, Lunt, Norry, Williams, Biswas, and Van~Tine]{Angelo2024TCRT}
Sandra~P D’Angelo, Dejka~M Araujo, Albiruni R~Abdul Razak, Mark Agulnik, Steven Attia, Jean-Yves Blay, Irene~Carrasco Garcia, John~A Charlson, Edwin Choy, George~D Demetri, Mihaela Druta, Edouard Forcade, Kristen~N Ganjoo, John Glod, Vicki~L Keedy, Axel~Le Cesne, David~A Liebner, Victor Moreno, Seth~M Pollack, Scott~M Schuetze, Gary~K Schwartz, Sandra~J Strauss, William~D Tap, Fiona Thistlethwaite, Claudia Maria~Valverde Morales, Michael~J Wagner, Breelyn~A Wilky, Cheryl McAlpine, Laura Hudson, Jean-Marc Navenot, Tianjiao Wang, Jane Bai, Stavros Rafail, Ruoxi Wang, Amy Sun, Lilliam Fernandes, Erin Van~Winkle, Erica Elefant, Colin Lunt, Elliot Norry, Dennis Williams, Swethajit Biswas, and Brian~A Van~Tine.
\newblock {Afamitresgene autoleucel for advanced synovial sarcoma and myxoid round cell liposarcoma (SPEARHEAD-1): an international, open-label, phase 2 trial}.
\newblock \emph{The Lancet}, 403\penalty0 (10435):\penalty0 1460--1471, 2024.
\newblock \doi{10.1016/S0140-6736(24)00319-2}.
\newblock URL \url{https://doi.org/10.1016/S0140-6736(24)00319-2}.

\bibitem[Feng et~al.(2024)Feng, Chen, Hai, Pang, Zheng, Xie, Zhang, Li, Zhang, Liu, et~al.]{feng2024sliding}
Ziyan Feng, Jingyang Chen, Youlong Hai, Xuelian Pang, Kun Zheng, Chenglong Xie, Xiujuan Zhang, Shengqing Li, Chengjuan Zhang, Kangdong Liu, et~al.
\newblock Sliding-attention transformer neural architecture for predicting t cell receptor--antigen--human leucocyte antigen binding.
\newblock \emph{Nature Machine Intelligence}, 6\penalty0 (10):\penalty0 1216--1230, 2024.

\bibitem[Gao et~al.(2024)Gao, Dong, Gao, Jin, Yang, Yan, and Liu]{gao2024unitcr}
Yicheng Gao, Kejing Dong, Yuli Gao, Xuan Jin, Jingya Yang, Gang Yan, and Qi~Liu.
\newblock {Unified cross-modality integration and analysis of T cell receptors and T cell transcriptomes by low-resource-aware representation learning}.
\newblock \emph{Cell Genomics}, 2024.
\newblock ISSN 2666-979X.
\newblock \doi{10.1016/j.xgen.2024.100553}.
\newblock URL \url{https://doi.org/10.1016/j.xgen.2024.100553}.

\bibitem[Gheini et~al.(2021)Gheini, Ren, and May]{gheini2021cross}
Mozhdeh Gheini, Xiang Ren, and Jonathan May.
\newblock Cross-attention is all you need: {A}dapting pretrained {T}ransformers for machine translation.
\newblock In \emph{{Proceedings of the 2021 Conference on Empirical Methods in Natural Language Processing}}, pp.\  1754--1765. Association for Computational Linguistics, 2021.
\newblock \doi{10.18653/v1/2021.emnlp-main.132}.
\newblock URL \url{https://aclanthology.org/2021.emnlp-main.132}.

\bibitem[Han et~al.(2023)Han, Yang, Tian, Fattah, Itzstein, Hu, Zhang, Kang, Yang, Liu, Xue, Liang, Raman, Zhu, Xiao, Dowell, Homsi, Rashdan, Yang, Gwin, Hsiehchen, Gloria-McCutchen, Pan, Wu, Gibbons, Wang, Yee, Huang, Reuben, Cheng, Zhang, Gerber, and Wang]{han2023pan}
Yi~Han, Yuqiu Yang, Yanhua Tian, Farjana~J Fattah, Mitchell S~von Itzstein, Yifei Hu, Minying Zhang, Xiongbin Kang, Donghan~M Yang, Jialiang Liu, Yaming Xue, Chaoying Liang, Indu Raman, Chengsong Zhu, Olivia Xiao, Jonathan~E Dowell, Jade Homsi, Sawsan Rashdan, Shengjie Yang, Mary~E Gwin, David Hsiehchen, Yvonne Gloria-McCutchen, Ke~Pan, Fangjiang Wu, Don Gibbons, Xinlei Wang, Cassian Yee, Junzhou Huang, Alexandre Reuben, Chao Cheng, Jianjun Zhang, David~E Gerber, and Tao Wang.
\newblock {pan-MHC and cross-Species Prediction of T Cell Receptor-Antigen Binding}.
\newblock \emph{bioRxiv}, 2023.
\newblock ISSN 2692-8205.
\newblock \doi{10.1101/2023.12.01.569599}.
\newblock URL \url{https://doi.org/10.1101/2023.12.01.569599}.

\bibitem[Hassel et~al.(2023)Hassel, Piperno-Neumann, Rutkowski, Baurain, Schlaak, Butler, Sullivan, Dummer, Kirkwood, Orloff, Sacco, Ochsenreither, Joshua, Gastaud, Curti, Piulats, Salama, Shoushtari, Demidov, Milhem, Chmielowski, Kim, Carvajal, Hamid, Collins, Ranade, Holland, Pfeiffer, and Nathan]{Haseel2023TCRTtrial}
Jessica~C Hassel, Sophie Piperno-Neumann, Piotr Rutkowski, Jean-Francois Baurain, Max Schlaak, Marcus~O Butler, Ryan~J Sullivan, Reinhard Dummer, John~M Kirkwood, Marlana Orloff, Joseph~J Sacco, Sebastian Ochsenreither, Anthony~M Joshua, Lauris Gastaud, Brendan Curti, Josep~M Piulats, April K~S Salama, Alexander~N Shoushtari, Lev Demidov, Mohammed Milhem, Bartosz Chmielowski, Kevin~B Kim, Richard~D Carvajal, Omid Hamid, Laura Collins, Koustubh Ranade, Chris Holland, Constance Pfeiffer, and Paul Nathan.
\newblock {Three-Year overall survival with Tebentafusp in metastatic uveal melanoma}.
\newblock \emph{New England Journal of Medicine}, 389\penalty0 (24):\penalty0 2256--2266, 2023.
\newblock \doi{10.1056/NEJMoa2304753}.
\newblock URL \url{https://doi.org/10.1056/NEJMoa2304753}.

\bibitem[Heitmann et~al.(2022)Heitmann, Bilich, Tandler, Nelde, Maringer, Marconato, Reusch, J{\"a}ger, Denk, Richter, Anton, Weber, Roerden, Bauer, Rieth, Wacker, H{\"o}rber, Peter, Meisner, Fischer, L{\"o}ffler, Karbach, J{\"a}ger, Klein, Rammensee, Salih, and Walz]{Heitmann2022TcellVaccine}
Jonas~S. Heitmann, Tatjana Bilich, Claudia Tandler, Annika Nelde, Yacine Maringer, Maddalena Marconato, Julia Reusch, Simon J{\"a}ger, Monika Denk, Marion Richter, Leonard Anton, Lisa~Marie Weber, Malte Roerden, Jens Bauer, Jonas Rieth, Marcel Wacker, Sebastian H{\"o}rber, Andreas Peter, Christoph Meisner, Imma Fischer, Markus~W. L{\"o}ffler, Julia Karbach, Elke J{\"a}ger, Reinhild Klein, Hans-Georg Rammensee, Helmut~R. Salih, and Juliane~S. Walz.
\newblock {A COVID-19 peptide vaccine for the induction of SARS-CoV-2 T cell immunity}.
\newblock \emph{Nature}, 601\penalty0 (7894):\penalty0 617--622, 2022.
\newblock ISSN 1476-4687.
\newblock \doi{10.1038/s41586-021-04232-5}.
\newblock URL \url{https://doi.org/10.1038/s41586-021-04232-5}.

\bibitem[Heitmann et~al.(2023)Heitmann, Tandler, Marconato, Nelde, Habibzada, Rittig, Tegeler, Maringer, Jaeger, Denk, Richter, Oezbek, Wiesm{\"u}ller, Bauer, Rieth, Wacker, Schroeder, Hoenisch~Gravel, Scheid, M{\"a}rklin, Henrich, Klimovich, Clar, Lutz, Holzmayer, H{\"o}rber, Peter, Meisner, Fischer, L{\"o}ffler, Peuker, Habringer, Goetze, J{\"a}ger, Rammensee, Salih, and Walz]{Heitmann2023NC}
Jonas~S. Heitmann, Claudia Tandler, Maddalena Marconato, Annika Nelde, Timorshah Habibzada, Susanne~M. Rittig, Christian~M. Tegeler, Yacine Maringer, Simon~U. Jaeger, Monika Denk, Marion Richter, Melek~T. Oezbek, Karl-Heinz Wiesm{\"u}ller, Jens Bauer, Jonas Rieth, Marcel Wacker, Sarah~M. Schroeder, Naomi Hoenisch~Gravel, Jonas Scheid, Melanie M{\"a}rklin, Annika Henrich, Boris Klimovich, Kim~L. Clar, Martina Lutz, Samuel Holzmayer, Sebastian H{\"o}rber, Andreas Peter, Christoph Meisner, Imma Fischer, Markus~W. L{\"o}ffler, Caroline~Anna Peuker, Stefan Habringer, Thorsten~O. Goetze, Elke J{\"a}ger, Hans-Georg Rammensee, Helmut~R. Salih, and Juliane~S. Walz.
\newblock {Phase I/II trial of a peptide-based COVID-19 T-cell activator in patients with B-cell deficiency}.
\newblock \emph{Nature Communications}, 14\penalty0 (1):\penalty0 5032, 2023.
\newblock ISSN 2041-1723.
\newblock \doi{10.1038/s41467-023-40758-0}.
\newblock URL \url{https://doi.org/10.1038/s41467-023-40758-0}.

\bibitem[Hou et~al.(2019)Hou, Chang, Ma, Shan, and Chen]{hou2019cross}
Ruibing Hou, Hong Chang, Bingpeng Ma, Shiguang Shan, and Xilin Chen.
\newblock Cross attention network for few-shot classification.
\newblock In \emph{Advances in Neural Information Processing Systems}, volume~32, 2019.
\newblock ISBN 9781713807933.

\bibitem[Huppa et~al.(2010)Huppa, Axmann, M{\"o}rtelmaier, Lillemeier, Newell, Brameshuber, Klein, Sch{\"u}tz, and Davis]{Huppa2010pMHCTCR}
Johannes~B Huppa, Markus Axmann, Manuel~A M{\"o}rtelmaier, Bj{\"o}rn~F Lillemeier, Evan~W Newell, Mario Brameshuber, Lawrence~O Klein, Gerhard~J Sch{\"u}tz, and Mark~M Davis.
\newblock {TCR--peptide--MHC interactions in situ show accelerated kinetics and increased affinity}.
\newblock \emph{Nature}, 463\penalty0 (7283):\penalty0 963--967, 2010.
\newblock ISSN 1476-4687.
\newblock \doi{10.1038/nature08746}.
\newblock URL \url{https://doi.org/10.1038/nature08746}.

\bibitem[Jin et~al.(2021)Jin, Liu, Nasiri, Cui, Louis, Zhang, Zhao, and Hu]{jin2021DeepAttentionPan}
Jing Jin, Zhonghao Liu, Alireza Nasiri, Yuxin Cui, Stephen-Yves Louis, Ansi Zhang, Yong Zhao, and Jianjun Hu.
\newblock Deep learning pan-specific model for interpretable mhc-i peptide binding prediction with improved attention mechanism.
\newblock \emph{Proteins: Structure, Function, and Bioinformatics}, 89\penalty0 (7):\penalty0 866--883, 2021.

\bibitem[Jin et~al.(2023)Jin, Wu, Chen, Pan, Wang, Xie, Quan, and Lyu]{jin2023capla}
Zhi Jin, Tingfang Wu, Taoning Chen, Deng Pan, Xuejiao Wang, Jingxin Xie, Lijun Quan, and Qiang Lyu.
\newblock {CAPLA: improved prediction of protein--ligand binding affinity by a deep learning approach based on a cross-attention mechanism}.
\newblock \emph{Bioinformatics}, 39\penalty0 (2):\penalty0 btad049, 2023.
\newblock ISSN 1367-4811.
\newblock \doi{10.1093/bioinformatics/btad049}.
\newblock URL \url{https://doi.org/10.1093/bioinformatics/btad049}.

\bibitem[Ju et~al.(2021)Ju, Zhang, Xiao, Li, Li, Zhang, and Zhou]{ju2021joint}
Xincheng Ju, Dong Zhang, Rong Xiao, Junhui Li, Shoushan Li, Min Zhang, and Guodong Zhou.
\newblock Joint multi-modal aspect-sentiment analysis with auxiliary cross-modal relation detection.
\newblock In \emph{Proceedings of the 2021 Conference on Empirical Methods in Natural Language Processing}, pp.\  4395--4405. Association for Computational Linguistics, 2021.
\newblock \doi{10.18653/v1/2021.emnlp-main.360}.
\newblock URL \url{https://aclanthology.org/2021.emnlp-main.360}.

\bibitem[Jumper et~al.(2021)Jumper, Evans, Pritzel, Green, Figurnov, Ronneberger, Tunyasuvunakool, Bates, {\v{Z}}{\'\i}dek, Potapenko, Bridgland, Meyer, Kohl, Ballard, Cowie, Romera-Paredes, Nikolov, Jain, Adler, Back, Petersen, Reiman, Clancy, Zielinski, Steinegger, Pacholska, Berghammer, Bodenstein, Silver, Vinyals, Senior, Kavukcuoglu, Kohli, and Hassabis]{jumper2021af2}
John Jumper, Richard Evans, Alexander Pritzel, Tim Green, Michael Figurnov, Olaf Ronneberger, Kathryn Tunyasuvunakool, Russ Bates, Augustin {\v{Z}}{\'\i}dek, Anna Potapenko, Alex Bridgland, Clemens Meyer, Simon A.~A. Kohl, Andrew~J. Ballard, Andrew Cowie, Bernardino Romera-Paredes, Stanislav Nikolov, Rishub Jain, Jonas Adler, Trevor Back, Stig Petersen, David Reiman, Ellen Clancy, Michal Zielinski, Martin Steinegger, Michalina Pacholska, Tamas Berghammer, Sebastian Bodenstein, David Silver, Oriol Vinyals, Andrew~W. Senior, Koray Kavukcuoglu, Pushmeet Kohli, and Demis Hassabis.
\newblock {Highly accurate protein structure prediction with AlphaFold}.
\newblock \emph{{Nature}}, 596\penalty0 (7873):\penalty0 583--589, 2021.
\newblock \doi{https://doi.org/10.1038/s41586-021-03819-2}.

\bibitem[Kalemati et~al.(2023)Kalemati, Darvishi, and Koohi]{kalemati2023capsnet}
Mahmood Kalemati, Saeid Darvishi, and Somayyeh Koohi.
\newblock Capsnet-mhc predicts peptide-mhc class i binding based on capsule neural networks.
\newblock \emph{Communications Biology}, 6\penalty0 (1):\penalty0 492, 2023.

\bibitem[Kammertoens \& Blankenstein(2013)Kammertoens and Blankenstein]{Kammertoens2013pMHC}
Thomas Kammertoens and Thomas Blankenstein.
\newblock {It’s the Peptide-MHC Affinity, Stupid}.
\newblock \emph{Cancer Cell}, 23\penalty0 (4):\penalty0 429--431, 2013.
\newblock ISSN 1535-6108.
\newblock \doi{https://doi.org/10.1016/j.ccr.2013.04.004}.
\newblock URL \url{https://www.sciencedirect.com/science/article/pii/S1535610813001384}.

\bibitem[Lin et~al.(2023)Lin, Akin, Rao, Hie, Zhu, Lu, Smetanin, Verkuil, Kabeli, Shmueli, et~al.]{lin2023evolutionary}
Zeming Lin, Halil Akin, Roshan Rao, Brian Hie, Zhongkai Zhu, Wenting Lu, Nikita Smetanin, Robert Verkuil, Ori Kabeli, Yaniv Shmueli, et~al.
\newblock Evolutionary-scale prediction of atomic-level protein structure with a language model.
\newblock \emph{Science}, 379\penalty0 (6637):\penalty0 1123--1130, 2023.

\bibitem[Liu et~al.(2022)Liu, Chen, Cao, Jia, Rui, Zheng, Huang, Liu, Liu, Zhao, Lu, and Lin]{Liu2022CMVTCRT}
Guangna Liu, Hua Chen, Xingyu Cao, Lemei Jia, Wei Rui, Hongli Zheng, Daosheng Huang, Fang Liu, Yue Liu, Xueqiang Zhao, Peihua Lu, and Xin Lin.
\newblock {Efficacy of pp65-specific TCR-T cell therapy in treating cytomegalovirus infection after hematopoietic stem cell transplantation}.
\newblock \emph{American Journal of Hematology}, 97\penalty0 (11):\penalty0 1453--1463, 2022.
\newblock ISSN 0361-8609.
\newblock \doi{10.1002/ajh.26708}.
\newblock URL \url{https://doi.org/10.1002/ajh.26708}.

\bibitem[Lu et~al.(2021)Lu, Zhang, Zhu, Wang, Jiang, Xiao, Bernatchez, Heymach, Gibbons, Wang, Xu, Reuben, and Wang]{lu2021pMTnet}
Tianshi Lu, Ze~Zhang, James Zhu, Yunguan Wang, Peixin Jiang, Xue Xiao, Chantale Bernatchez, John~V Heymach, Don~L Gibbons, Jun Wang, Lin Xu, Alexandre Reuben, and Tao Wang.
\newblock {Deep learning-based prediction of the T cell receptor--antigen binding specificity}.
\newblock \emph{Nature Machine Intelligence}, 3\penalty0 (10):\penalty0 864--875, 2021.
\newblock ISSN 2522-5839.
\newblock \doi{10.1038/s42256-021-00383-2}.
\newblock URL \url{https://doi.org/10.1038/s42256-021-00383-2}.

\bibitem[Montemurro et~al.(2021)Montemurro, Schuster, Povlsen, Bentzen, Jurtz, Chronister, Crinklaw, Hadrup, Winther, Peters, et~al.]{montemurro2021nettcr}
Alessandro Montemurro, Viktoria Schuster, Helle~Rus Povlsen, Amalie~Kai Bentzen, Vanessa Jurtz, William~D Chronister, Austin Crinklaw, Sine~R Hadrup, Ole Winther, Bjoern Peters, et~al.
\newblock {NetTCR-2.0 enables accurate prediction of TCR-peptide binding by using paired TCR$\alpha$ and $\beta$ sequence data}.
\newblock \emph{Communications Biology}, 4\penalty0 (1):\penalty0 1060, 2021.
\newblock ISSN 2399-3642.
\newblock \doi{10.1038/s42003-021-02610-3}.
\newblock URL \url{https://doi.org/10.1038/s42003-021-02610-3}.

\bibitem[Mora \& Walczak(2019)Mora and Walczak]{Mora2019Diveristy}
Thierry Mora and Aleksandra~M Walczak.
\newblock Quantifying lymphocyte receptor diversity.
\newblock In Jayajit Das and Ciriyam Jayaprakash (eds.), \emph{{Systems Immunology: An Introduction to Modeling Methods for Scientists}}, pp.\  183--198. CRC Press, Taylor {\&} Francis Group, Boca Raton FL, United States, 2019.
\newblock ISBN 978-1-4987-1740-3.
\newblock \doi{10.1201/9781315119847}.

\bibitem[Parker et~al.(1994)Parker, Bednarek, and Coligan]{parker1994PSSM}
Kenneth~C Parker, Maria~A Bednarek, and John~E Coligan.
\newblock {Scheme for ranking potential HLA-A2 binding peptides based on independent binding of individual peptide side-chains.}
\newblock \emph{{The Journal of Immunology}}, 152\penalty0 (1):\penalty0 163--175, 1994.

\bibitem[Reynisson et~al.(2020)Reynisson, Alvarez, Paul, Peters, and Nielsen]{reynisson2020netmhcpan}
Birkir Reynisson, Bruno Alvarez, Sinu Paul, Bjoern Peters, and Morten Nielsen.
\newblock {NetMHCpan-4.1 and NetMHCIIpan-4.0: improved predictions of MHC antigen presentation by concurrent motif deconvolution and integration of MS MHC eluted ligand data}.
\newblock \emph{Nucleic Acids Research}, 48\penalty0 (W1):\penalty0 W449--W454, 2020.
\newblock ISSN 1362-4962.
\newblock \doi{10.1093/nar/gkaa379}.
\newblock URL \url{https://doi.org/10.1093/nar/gkaa379}.

\bibitem[Rock et~al.(2016)Rock, Reits, and Neefjes]{Rock2016MHCdiversity}
Kenneth~L Rock, Eric Reits, and Jacques Neefjes.
\newblock {Present Yourself! By MHC Class I and MHC Class II Molecules}.
\newblock \emph{Trends in Immunology}, 37\penalty0 (11):\penalty0 724--737, 2016.
\newblock ISSN 1471-4906.
\newblock \doi{10.1016/j.it.2016.08.010}.
\newblock URL \url{https://doi.org/10.1016/j.it.2016.08.010}.

\bibitem[Rojas et~al.(2023)Rojas, Sethna, Soares, Olcese, Pang, Patterson, Lihm, Ceglia, Guasp, Chu, Yu, Chandra, Waters, Ruan, Amisaki, Zebboudj, Odgerel, Payne, Derhovanessian, M{\"u}ller, Rhee, Yadav, Dobrin, Sadelain, {\L}uksza, Cohen, Tang, Basturk, G{\"o}nen, Katz, Do, Epstein, Momtaz, Park, Sugarman, Varghese, Won, Desai, Wei, D'Angelica, Kingham, Mellman, Merghoub, Wolchok, Sahin, T{\"u}reci, Greenbaum, Jarnagin, Drebin, O'Reilly, and Balachandran]{Rojas2023NeoAgtrial}
Luis~A Rojas, Zachary Sethna, Kevin~C Soares, Cristina Olcese, Nan Pang, Erin Patterson, Jayon Lihm, Nicholas Ceglia, Pablo Guasp, Alexander Chu, Rebecca Yu, Adrienne~Kaya Chandra, Theresa Waters, Jennifer Ruan, Masataka Amisaki, Abderezak Zebboudj, Zagaa Odgerel, George Payne, Evelyna Derhovanessian, Felicitas M{\"u}ller, Ina Rhee, Mahesh Yadav, Anton Dobrin, Michel Sadelain, Marta {\L}uksza, Noah Cohen, Laura Tang, Olca Basturk, Mithat G{\"o}nen, Seth Katz, Richard~Kinh Do, Andrew~S Epstein, Parisa Momtaz, Wungki Park, Ryan Sugarman, Anna~M Varghese, Elizabeth Won, Avni Desai, Alice~C Wei, Michael~I D'Angelica, T~Peter Kingham, Ira Mellman, Taha Merghoub, Jedd~D Wolchok, Ugur Sahin, {\"O}zlem T{\"u}reci, Benjamin~D Greenbaum, William~R Jarnagin, Jeffrey Drebin, Eileen~M O'Reilly, and Vinod~P Balachandran.
\newblock {Personalized RNA neoantigen vaccines stimulate T cells in pancreatic cancer}.
\newblock \emph{Nature}, 618\penalty0 (7963):\penalty0 144--150, 2023.
\newblock ISSN 1476-4687.
\newblock \doi{10.1038/s41586-023-06063-y}.
\newblock URL \url{https://doi.org/10.1038/s41586-023-06063-y}.

\bibitem[Springer et~al.(2021)Springer, Tickotsky, and Louzoun]{springer2021ergo2}
Ido Springer, Nili Tickotsky, and Yoram Louzoun.
\newblock Contribution of t cell receptor alpha and beta {CDR3}, {MHC} typing, {V} and {J} genes to peptide binding prediction.
\newblock \emph{Frontiers in Immunology}, 12:\penalty0 664514, 2021.
\newblock \doi{10.3389/fimmu.2021.664514}.

\bibitem[Sun et~al.(2021)Sun, Liu, Tao, and Lian]{sun2021multimodal}
Licai Sun, Bin Liu, Jianhua Tao, and Zheng Lian.
\newblock Multimodal cross-and self-attention network for speech emotion recognition.
\newblock In \emph{ICASSP 2021-2021 IEEE International Conference on Acoustics, Speech and Signal Processing (ICASSP)}, pp.\  4275--4279. IEEE, 2021.
\newblock \doi{10.1109/ICASSP39728.2021.9414654}.
\newblock URL \url{https://doi.org/10.1109/ICASSP39728.2021.9414654}.

\bibitem[Touvron et~al.(2021)Touvron, Cord, Sablayrolles, Synnaeve, and J{\'e}gou]{touvron2021going}
Hugo Touvron, Matthieu Cord, Alexandre Sablayrolles, Gabriel Synnaeve, and Herv{\'e} J{\'e}gou.
\newblock Going deeper with image transformers.
\newblock \emph{arXiv preprint arXiv:2103.17239}, 2021.
\newblock URL \url{https://arxiv.org/abs/2103.17239}.

\bibitem[Vaswani et~al.(2017)Vaswani, Shazeer, Parmar, Uszkoreit, Jones, Gomez, Kaiser, and Polosukhin]{vaswani2017attention}
Ashish Vaswani, Noam Shazeer, Niki Parmar, Jakob Uszkoreit, Llion Jones, Aidan~N Gomez, {\L}ukasz Kaiser, and Illia Polosukhin.
\newblock Attention is all you need.
\newblock In \emph{Advances in Neural Information Processing Systems}, volume~30, 2017.
\newblock ISBN 9781510860964.

\bibitem[Weber et~al.(2024)Weber, Carlino, Khattak, Meniawy, Ansstas, Taylor, Kim, McKean, Long, Sullivan, Faries, Tran, Cowey, Pecora, Shaheen, Segar, Medina, Atkinson, Gibney, Luke, Thomas, Buchbinder, Healy, Huang, Morrissey, Feldman, Sehgal, Robert-Tissot, Hou, Zhu, Brown, Aanur, Meehan, and Zaks]{Weber2024NeoAgTrial}
Jeffrey~S. Weber, Matteo~S. Carlino, Adnan Khattak, Tarek Meniawy, George Ansstas, Matthew~H. Taylor, Kevin~B. Kim, Meredith McKean, Georgina~V. Long, Ryan~J. Sullivan, Mark Faries, Thuy~T. Tran, C.~Lance Cowey, Andrew Pecora, Montaser Shaheen, Jennifer Segar, Theresa Medina, Victoria Atkinson, Geoffrey~T. Gibney, Jason~J. Luke, Sajeve Thomas, Elizabeth~I. Buchbinder, Jane~A. Healy, Mo~Huang, Manju Morrissey, Igor Feldman, Vasudha Sehgal, Celine Robert-Tissot, Peijie Hou, Lili Zhu, Michelle Brown, Praveen Aanur, Robert~S. Meehan, and Tal Zaks.
\newblock {Individualised neoantigen therapy mRNA-4157 (V940) plus pembrolizumab versus pembrolizumab monotherapy in resected melanoma (KEYNOTE-942): a randomised, phase 2b study}.
\newblock \emph{The Lancet}, 403\penalty0 (10427):\penalty0 632--644, 2024.
\newblock ISSN 0140-6736.
\newblock \doi{10.1016/S0140-6736(23)02268-7}.
\newblock URL \url{https://doi.org/10.1016/S0140-6736(23)02268-7}.

\bibitem[Wu et~al.(2023)Wu, Cao, Wu, Wu, Wang, and Duan]{wu2023ccbhla}
Yejian Wu, Lujing Cao, Zhipeng Wu, Xinyi Wu, Xinqiao Wang, and Hongliang Duan.
\newblock {CcBHLA: pan-specific peptide-HLA class I binding prediction via Convolutional and BiLSTM features}.
\newblock \emph{bioRxiv}, 2023.
\newblock ISSN 2692-8205.
\newblock \doi{10.1101/2023.04.24.538196}.
\newblock URL \url{https://doi.org/10.1101/2023.04.24.538196}.

\bibitem[Yang et~al.(2018)Yang, Wu, Bedbrook, and Arnold]{yang2018learned}
Kevin~K Yang, Zachary Wu, Claire~N Bedbrook, and Frances~H Arnold.
\newblock Learned protein embeddings for machine learning.
\newblock \emph{Bioinformatics}, 34\penalty0 (15):\penalty0 2642--2648, 2018.

\bibitem[Yarchoan et~al.(2024)Yarchoan, Gane, Marron, Perales-Linares, Yan, Cooch, Shu, Fertig, Kagohara, Bartha, Northcott, Lyle, Rochestie, Peters, Connor, Jaffee, Csiki, Weiner, Perales-Puchalt, and Sardesai]{Yarchoan2024NeoAgtrial}
Mark Yarchoan, Edward~J Gane, Thomas~U Marron, Renzo Perales-Linares, Jian Yan, Neil Cooch, Daniel~H Shu, Elana~J Fertig, Luciane~T Kagohara, Gabor Bartha, Josette Northcott, John Lyle, Sarah Rochestie, Joann Peters, Jason~T Connor, Elizabeth~M Jaffee, Ildiko Csiki, David~B Weiner, Alfredo Perales-Puchalt, and Niranjan~Y Sardesai.
\newblock Personalized neoantigen vaccine and pembrolizumab in advanced hepatocellular carcinoma: a phase 1/2 trial.
\newblock \emph{Nature Medicine}, 30\penalty0 (4):\penalty0 1044--1053, 2024.
\newblock ISSN 1546-170X.
\newblock \doi{10.1038/s41591-024-02894-y}.
\newblock URL \url{https://doi.org/10.1038/s41591-024-02894-y}.

\bibitem[Ye et~al.(2023)Ye, Li, Mi, Shao, Dai, Ding, Feng, Sun, Shen, and Xiao]{ye2023stmhcpan}
Zheng Ye, Shaohao Li, Xue Mi, Baoyi Shao, Zhu Dai, Bo~Ding, Songwei Feng, Bo~Sun, Yang Shen, and Zhongdang Xiao.
\newblock {STMHCpan, an accurate Star-Transformer-based extensible framework for predicting MHC I allele binding peptides}.
\newblock \emph{Briefings in Bioinformatics}, 24\penalty0 (3):\penalty0 bbad164, 2023.
\newblock ISSN 1477-4054.
\newblock \doi{10.1093/bib/bbad164}.
\newblock URL \url{https://doi.org/10.1093/bib/bbad164}.

\bibitem[Yu et~al.(2025)Yu, Fang, Tian, and Liu]{yu2025unifyimmun}
Chenpeng Yu, Xing Fang, Shiye Tian, and Hui Liu.
\newblock A unified cross-attention model for predicting antigen binding specificity to both hla and tcr molecules.
\newblock \emph{Nature Machine Intelligence}, 7:\penalty0 278--292, 2025.
\newblock \doi{10.1038/s42256-024-00973-w}.

\bibitem[Zhang et~al.(2009)Zhang, Lund, and Nielsen]{zhang2009pickpocket}
Hao Zhang, Ole Lund, and Morten Nielsen.
\newblock The pickpocket method for predicting binding specificities for receptors based on receptor pocket similarities: application to mhc-peptide binding.
\newblock \emph{Bioinformatics}, 25\penalty0 (10):\penalty0 1293--1299, 2009.

\bibitem[Zhang et~al.(2023)Zhang, Xie, Ding, and Wang]{zhang2023cross}
Jing Zhang, Yingshuai Xie, Weichao Ding, and Zhe Wang.
\newblock Cross on cross attention: Deep fusion transformer for image captioning.
\newblock \emph{IEEE Transactions on Circuits and Systems for Video Technology}, 33\penalty0 (8):\penalty0 4257--4268, 2023.
\newblock ISSN 1558-2205.
\newblock \doi{10.1109/TCSVT.2023.3243725}.
\newblock URL \url{https://doi.org/10.1109/TCSVT.2023.3243725}.

\bibitem[Zhao et~al.(2023)Zhao, He, Xu, Li, Xu, Su, He, Huang, Rossjohn, Song, et~al.]{zhao2023deepair}
Yu~Zhao, Bing He, Fan Xu, Chen Li, Zhimeng Xu, Xiaona Su, Haohuai He, Yueshan Huang, Jamie Rossjohn, Jiangning Song, et~al.
\newblock {DeepAIR: A deep learning framework for effective integration of sequence and 3D structure to enable adaptive immune receptor analysis}.
\newblock \emph{Science Advances}, 9\penalty0 (32):\penalty0 eabo5128, 2023.
\newblock ISSN 2375-2548.
\newblock \doi{10.1126/sciadv.abo5128}.
\newblock URL \url{https://doi.org/10.1126/sciadv.abo5128}.

\bibitem[Zhao et~al.(2025)Zhao, Yu, Su, Shu, Ma, Wang, Jiang, Wei, Li, Huang, Cheng, Ren, and Feng]{zhao2025unipmt}
Yunxiang Zhao, Jijun Yu, Yixin Su, You Shu, Enhao Ma, Jing Wang, Shuyang Jiang, Congwen Wei, Dongsheng Li, Zhen Huang, Gong Cheng, Hongguang Ren, and Jiannan Feng.
\newblock A unified deep framework for peptide--major histocompatibility complex--t cell receptor binding prediction.
\newblock \emph{Nature Machine Intelligence}, 7:\penalty0 650--660, 2025.
\newblock \doi{10.1038/s42256-025-01002-0}.

\end{thebibliography}

\newpage
\appendix
\crefalias{section}{appendix}
\crefalias{subsection}{subappendix}

\section{Implementation Details}
\label{app:training-details}

The complete models used AdamW with binary cross-entropy on logits, one epoch of linear warmup followed by cosine decay to 5\% of the initial learning rate over the planned training steps. Stage~2 transferred the same-seed Stage~1 weights, froze the transferred modules for the first ten epochs and then used 0.3 times the new-module learning rate for those modules. Dropout was 0.1. Table~\ref{tab:training_current} gives the settings recorded for all five complete-model runs. Baseline and ablation runs follow their corresponding model-specific protocols.

\noindent For the architecture of Fusion-pMT, we set the embedding dimension to 64, use 4 attention heads, and use 256-dimensional hidden layers in the feed-forward networks (FFNs). The instantiated Fusion-pM and Fusion-pMT models contain 252,033 and 452,546 parameters in total, respectively (\Cref{tab:training_current}). Both models use a single linear classification layer.\par

\subsection{Attention Blocks and Normalization}
\textbf{Activations and Normalization.} The molecule-specific sequence encoders and cross-attention blocks use post-residual LayerNorm and ReLU activations in their FFNs. The stage-level self-attention refinement blocks use Pre-LN residual updates with LayerScale and GELU activations. The class-attention readouts also use Pre-LN updates, LayerScale, and GELU, followed by an output LayerNorm. Residual connections are applied around the attention and FFN sublayers.
 
In the class-attention readout, the first residual bypass carries the unnormalized CLS token, and the second carries its post-attention representation. Figure~\ref{fig:Model_Structure}(d) displays token-wise LayerNorm before concatenation. Applying this shared normalization after concatenation gives the same result. Dropout, padding masks, activations and internal attention projections are specified by the implementation and omitted from the schematic for clarity.
\par
\label{sec:imple-details}

\section{Protein Sequence Embedding}

This appendix describes the residue features and positional encoding
used in our models.

\label{app:seq-tech}

\subsection{Shared Sequence Encoder} 
\label{app:protein-seq}

We start with the embedding of protein sequences in this subsection,
and then introduce the positional encoding method we adopt in \Cref{app:pe}.

Following the notation in \Cref{sec:ca-in-bio}, let
\(\mathbf{X}=[a_1,a_2,\ldots,a_L]\) denote an input sequence
of \(L\) residues.

Here, each $a_i$ indicates a certain amino acid, and {the model alphabet contains the 20 standard amino acids plus the unknown-residue symbol $\mathtt{X}$ and is denoted as} 
$$\mathcal{A} \defeq \{\mathtt{A, C, D, E, F, G, H, I, K, L, M, N, P, Q, R, S, T, V, W, X, Y}\}$$
in terms of letter abbreviations of amino acids.

\textbf{One-hot encoding.}
Each residue is represented by its amino-acid identity. Here, each amino acid \( a_i \) corresponds to a one-hot encoded vector \( \mathbf{h}(a_i) \) of length \( |\mathcal{A}| \):
$$\mathbf{h}(a_i) = [h_1, h_2, \ldots, h_{|\mathcal{A}|}]^{\mathsf T},$$
where \( h_j = 1 \) if \( a_i \) is the \( j \)-th amino acid in \( \mathcal{A} \), and \( h_j = 0 \) otherwise.

The one-hot encoding of the entire sequence \( \mathbf{X} \) is
\[
   \mathbf{O}(\mathbf{X})
   = [\mathbf{h}(a_1), \mathbf{h}(a_2), \dots,
      \mathbf{h}(a_L)]^{\mathsf T}
   \in \{0,1\}^{L \times |\mathcal{A}|}.
\]

\textbf{Residue features and shared projection.} For each residue, the 21-dimensional one-hot vector is concatenated with four binary physicochemical indicators: aromatic (F, W, Y), acidic (D, E), basic (K, R, H), and hydrophilic (N, Q, S, T, Y). These indicators are defined as 0/1 values in the model implementation, with no additional normalization. The resulting 25-dimensional vector is mapped to 64 dimensions by one bias-free linear projection shared by the peptide, MHC, and TCR inputs. The unknown symbol $\mathtt{X}$ has its own one-hot entry and four zero-valued indicators. Padding uses a separate token with zero-valued input features; prepared sequences are padded to the model's 35-position internal width, and padding is masked as keys and values in attention and in the class-attention readouts.

\textbf{Contextual encoding and readout.} After the shared projection, sinusoidal positional encodings are added as described in \Cref{app:pe}. Each molecular sequence then passes through its own Self-Attention and feed-forward network (FFN) block, with independent parameters. The residue-level token sequences are retained for the Stage 1 peptide--MHC and Stage 2 pMHC--TCR cross-attention modules; sequence summaries are produced by the subsequent class-attention readouts, as detailed in \Cref{sec:readout-prediction}. \par

\subsection{Positional Encoding for Protein Sequences}
\label{app:pe}

The positional encoding matrix $\mathbf{P}$ in
\Cref{eq:residue-representation} records absolute residue positions.
Let $p(s,j)$ denote its entry at sequence position $s$ and
feature dimension $j=0,\ldots,d_{\mathrm{model}}-1$. Following \citet{vaswani2017attention},
for $i=0,\ldots,d_{\mathrm{model}}/2-1$, with
$d_{\mathrm{model}}=64$, we use:

\begin{align*}
p(s, 2i) &= \sin\left(\frac{s}{10000^{2i/d_{\mathrm{model}}}}\right), \\
p(s, 2i+1) &= \cos\left(\frac{s}{10000^{2i/d_{\mathrm{model}}}}\right),
\end{align*}
Positional encoding identifies residue positions within each sequence,
{and we add it to the 64-dimensional projected residue tokens before each molecule's Self-Attention/FFN block. Positions are indexed from $s=0$ independently for the peptide, MHC, and TCR sequences; the same deterministic, non-learned sinusoidal formula produces the same positional values at corresponding positions in all three streams. The Self-Attention/FFN parameters remain independent, and the cross-attention modules operate on token representations that already contain position information.}

\section{Useful Facts}
\label{app:bio-facts}

\subsection{Biological Molecules of Adaptive Immunity}
\label{app:immune-bio-mol}

\begin{figure}[H]
    \centering
    \includegraphics[width=0.42\linewidth]{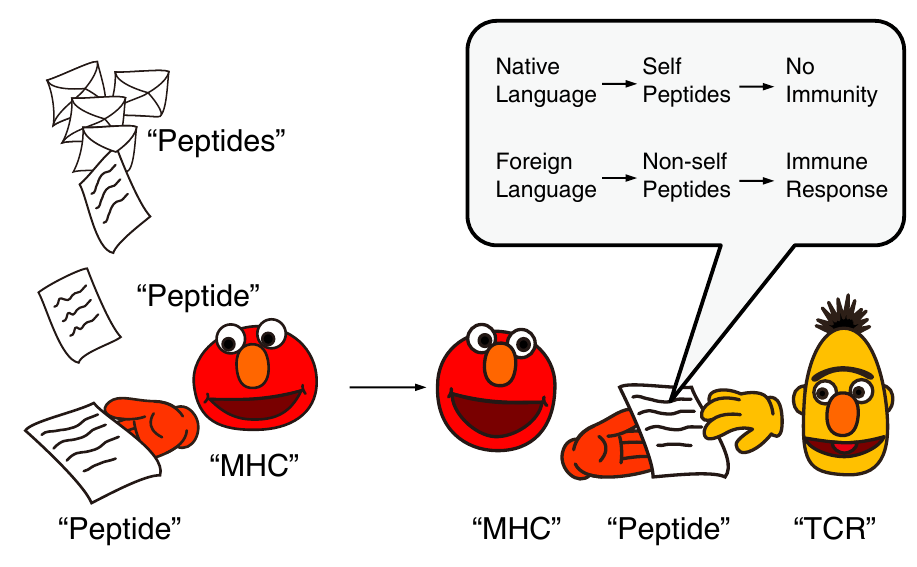}
    \caption{\textbf{Illustrative self and foreign peptide examples.} Recognition and immune activation depend on receptor specificity and immune regulation.}
    \label{fig:Immu-Language}
\end{figure}

In adaptive immunity, the major players are the highly diverse B and T cells, with unique surface receptors known as B cell receptors (BCRs) and T cell receptors (TCRs), respectively. These cells recognize specific parts of an antigen, referred to as epitopes. However, the mechanisms of antigen recognition differ between B and T cells. B cells target a fragment of the antigen known as a B cell epitope. Recognition by BCRs primarily depends on three-dimensional conformational information from the fragment, which contains mainly non-contiguous amino acid residues. On the other hand, T cell epitopes, recognized by TCRs, depend on their binding to major histocompatibility complex (MHC) molecules. These epitopes are linear, formed by contiguous amino acid residues.

\textbf{Major Histocompatibility Complex (MHC).}
The \textbf{major histocompatibility complex (MHC)} molecules are cell-surface proteins essential for adaptive immunity. In humans, MHC genes are called human leukocyte antigens (\textit{HLA}s). The MHC class I molecules present endogenous peptides from proteins self-generated intracellularly, while the MHC class II molecules are mainly expressed on antigen presenting cells. The MHC class I molecules contain an $\alpha$ chain from \textit{MHC} class I genes and $\beta_{2}$ microglobulin ($\beta_{2}m$), which can present peptides  ranging from 8 to 12 amino acids.  MHC class II molecules consist of one $\alpha$ and one $\beta$ chain, allowing the binding of longer peptides ranging from 9 to 25 amino acid residues, or even longer. The peptide repertoires presented by MHC class I and class II molecules are highly diverse and are shaped by HLA polymorphism and allele-specific peptide-binding preferences~\citep{Rock2016MHCdiversity}.

\textbf{T Cell Receptor (TCR).}
The \textbf{T cell receptor (TCR)} is a protein complex on the T-cell surface that recognizes antigenic peptides presented by MHC molecules. A conventional $\alpha\beta$ TCR contains an $\alpha$ chain encoded at the \textit{TRA} locus and a $\beta$ chain encoded at the \textit{TRB} locus. TCR diversity arises from V(D)J recombination and junctional nucleotide addition and deletion, followed by thymic selection. Estimates of theoretical sequence space, single-chain diversity, paired $\alpha\beta$ diversity, and the number of clonotypes realized in an individual describe different quantities and should not be compared directly~\citep{Mora2019Diveristy,Chi2024cell}. Because the datasets used in this study primarily contain TCR $\beta$-chain CDR3 sequences, our data-driven modeling focuses on the $\beta$ chain.

\textbf{B Cell Receptor (BCR).}
{The \textbf{B cell receptor (BCR)} is a membrane-bound complex comprising an antigen-binding immunoglobulin and the Ig$\alpha$/Ig$\beta$ (CD79A/CD79B) signaling subunits. The immunoglobulin contains two heavy and two light chains. An antibody, a secreted form of immunoglobulin produced by a B cell, shares an identical antigen-binding region sequence with the B cell receptor (BCR) on the surface of that same cell~\citep{dong2022bcrassembly}. BCR diversity arises from V(D)J recombination and can be further altered by somatic hypermutation. The approximately $10^{21}$ estimate reported by \citet{Mora2019Diveristy} describes the generation-process Shannon diversity of heavy-chain sequences, not the number of possible complete BCR rearrangements.}

\begin{table}[t!]
\centering
\small
\caption{Overview of Molecules Involved in Antigen Presentation.}
\label{tab:antigen_presentation}
\resizebox{\textwidth}{!}{
\begin{tabular}{lccccc}
\toprule
\textbf{Molecule} & \textbf{Location} & \textbf{Total Length (aa)} & \textbf{Active Region} & \textbf{Main Function} & \textbf{Homology} \\
\midrule

MHC Class I & Cell Surface 
& \makecell{$\alpha$: $\sim$360 \\ $\beta_2m$: $\sim$120} 
& \makecell{Relevant:\\ $\alpha$ chain} 
& Present peptides to $CD8^{+}$ T cells 
& Varies \\
\hline

MHC Class II & Cell Surface 
& \makecell{$\alpha$ \& $\beta$: \\ 260--280} 
& \makecell{Relevant:\\ $\alpha_1$ and $\beta_1$ domains} 
& Present peptides to $CD4^{+}$ T cells 
& Varies \\
\hline

T Cell Receptor & T Cell Surface 
& \makecell{$\alpha$: 223 \\ $\beta$: 247} 
& \makecell{Variable regions:\\ 110--120 each chain} 
& Recognize peptide--MHC complexes 
& Low \\
\hline

BCR (mIg component) 
& \makecell{B Cell Surface} 
& \makecell{Light: 211--217 \\ Heavy: $\sim$450/$\sim$550} 
& \makecell{Variable domain:\\ 110} 
& Recognize antigens 
& Low \\

\bottomrule
\end{tabular}
}
\end{table}

\subsection{Protein Interactions and Prediction Methods}
\label{protein inter}
\label{app:related-methods}

Table~\ref{tab:related} summarizes how existing models represent
molecular inputs and combine their features. Fusion-pMT retains
residue-level representations and uses two-stage cross-attention
to connect peptide--MHC modeling with TCR interaction prediction.

\begin{table*}[t]
    \centering
    \small
    \caption{Comparison of immunological sequence models and their interaction or feature-fusion mechanisms. ESM-2 + MLP denotes our downstream baseline; DeepAIR fuses receptor gene, sequence, and structure features for TCRs and BCRs.
    }
    
    \setlength{\tabcolsep}{3pt}
    \renewcommand{\arraystretch}{1.15}
    \begin{tabular}{@{}
        p{0.18\textwidth}
        p{0.18\textwidth}
        p{0.17\textwidth}
        p{0.16\textwidth}
        p{0.23\textwidth}@{}}
        \toprule
        \textbf{Model} & \textbf{MHC} & \textbf{Peptide} & \textbf{TCR} & \textbf{Interaction / Fusion} \\
        \midrule
        STMHCpan~(\citeyear{ye2023stmhcpan}) & 34-residue MHC-I pseudosequence & Peptide sequence & N/A & Star-Transformer\\
        
        TransPHLA~(\citeyear{chu2021transmut,chu2022transphla}) & Self-Attention & Self-Attention & N/A & Joint Self-Attention \\
        
        PISTE~(\citeyear{feng2024sliding})& 1D CNN & 1D CNN & 1D CNN & Sliding-Attention\\
        
        NetMHCpan~(\citeyear{reynisson2020netmhcpan}) & ANN & ANN & N/A & Joint input \\
        
        CcBHLA~(\citeyear{wu2023ccbhla}) & BiLSTM & BiLSTM & N/A & CNN \\

        ESM-2 + MLP~(\citeyear{lin2023evolutionary}) & Self-Attention & Self-Attention & Self-Attention & Concat + MLP \\
        
        UniTCR~(\citeyear{gao2024unitcr}) & N/A & 2-layer LSTM & Self-Attention & Epitope--TCR Cross-Attention \\
        
        DeepAIR~(\citeyear{zhao2023deepair}) & N/A & N/A & Self-Attention & Gated attention + tensor fusion \\
        
        NetTCR~(\citeyear{montemurro2021nettcr}) & N/A & 1D CNN & 1D CNN & Concat \\
        
        pMTnet~(\citeyear{lu2021pMTnet}) & Joint LSTM & Joint LSTM & Autoencoder & Concat \\
        
        \textbf{Fusion-pMT (This work)} & \textbf{Shared proj. + Self-Attn} & \textbf{Shared proj. + Self-Attn} & \textbf{Shared proj. + Self-Attn} & \textbf{Two-stage Cross-Attention} \\
        \bottomrule
    \end{tabular}
    
    \label{tab:related}
\end{table*}

\section{Applications, Limitations, and Asset Licenses}

\subsection{Broader Impact and Limitations}
\label{app:limitations}

\noindent\textbf{Evaluation scope.} Our evaluation uses fixed public benchmark datasets with constructed negative examples, so the reported performance depends on their labeling and sampling protocols. Future work should evaluate robustness to unseen peptides, MHC types, or receptor distributions. The model uses TCR CDR3$\beta$ sequences and does not incorporate paired $\alpha\beta$ receptor information. Variation across training seeds reflects training variability on fixed test examples, rather than uncertainty across biological populations. Predicted binding scores also do not directly establish functional T-cell activation or clinical efficacy. Evaluation under distinct sampling protocols and experimental validation would be needed to assess these broader applications.
\noindent\textbf{Data representativeness and application risks.} Public immunological databases such as IEDB and VDJdb may provide uneven coverage of MHC alleles, antigens, and study populations. Aggregate benchmark scores therefore do not establish population-specific performance. Before predictions inform healthcare applications, validation should examine performance across relevant MHC groups and intended populations, including the consequences of false-positive and false-negative predictions. These assessments are needed to identify disparities that aggregate metrics could obscure.

\subsection{Impacts in Healthcare}
\label{app:impacts-healthcare}

\begin{figure}[t!]
  \centering
  \includegraphics[width=0.7\linewidth]{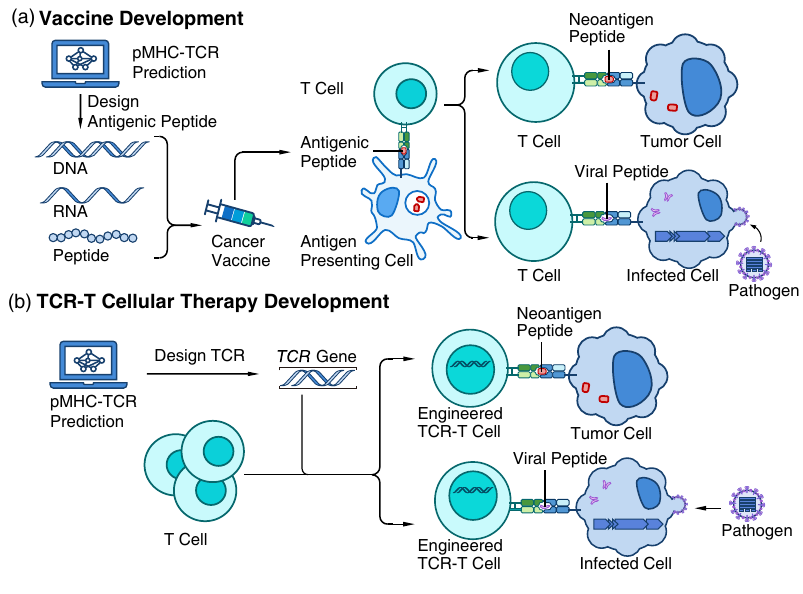}  \caption{
  \textbf{The Application of pMHC--TCR Binding Prediction in Healthcare}
  \textbf{(a)} Schematic Representation of the Therapeutic Cancer Vaccine.  \textbf{(b)} Schematic Representation of the Engineered TCR-T Cell Therapy. }
\label{fig:Therapy}
\end{figure}

In healthcare, the prediction of pMHC--TCR interactions has significant implications in the development of advanced therapies against cancers or infectious diseases (Figure~\ref{fig:Therapy}).

\textbf{Vaccine Development.} Understanding which peptides can bind to MHC molecules and be recognized by TCRs can help in the design of more effective vaccines, especially the neoantigen-based cancer vaccine. Neoantigens are newly generated peptides from somatic mutations that can be recognized by TCRs of tumor-specific T cells. Once mutations have been identified from matched tumor-normal sequencing data, candidate neoantigens can be computationally predicted and ranked by their predicted ability to stimulate a T cell response. A neoantigen-based cancer vaccine has shown promising results in a phase IIb study \citep{Weber2024NeoAgTrial}. Predicting pMHC--TCR interactions may help prioritize candidate neoantigens for experimental evaluation \citep{Rojas2023NeoAgtrial,Yarchoan2024NeoAgtrial} (Figure~\ref{fig:Therapy}(a)). 

In addition to cancers, pMHC--TCR prediction can also accelerate the development of infectious disease vaccines. During the COVID-19 pandemic, T-cell-directed vaccines have been designed in the form of peptides \citep{Heitmann2022TcellVaccine} and mRNA \citep{Arieta2023BNT162b4}. More precise prediction of pMHC--TCR interactions can improve the development of T-cell-directed vaccines in patients with immunodeficiency in phase I/II study \citep{Heitmann2023NC}.

\textbf{TCR-T Cellular Therapy.} The prediction of pMHC--TCR interactions can also aid in the development of T cell therapies, where the goal is to enhance the immune system’s ability to recognize and destroy abnormal cells. {In autologous TCR-engineered T-cell therapy, a tumor-antigen-specific TCR is introduced into a patient's T cells. The engineered cells are expanded and infused back into the patient~\citep{Angelo2024TCRT}. Afamitresgene autoleucel (Tecelra), an autologous TCR T-cell therapy, received accelerated approval from the U.S. Food and Drug Administration (FDA) on August 1, 2024 (\href{https://www.fda.gov/media/180568/download}{FDA approval letter}).} Additionally, the proof of concept for using TCR-T therapies against infectious diseases has been demonstrated in the treatment of cytomegalovirus infection after hematopoietic stem cell transplantation \citep{Liu2022CMVTCRT}, which sheds light on the broader application of TCR-T cellular therapies (Figure~\ref{fig:Therapy}(b)).

\subsection{Asset Licenses}
\label{app:license}

\begin{itemize}[itemsep=-0.2em, topsep=0.0em, leftmargin=1.0em]
    \item \textbf{TransPHLA-AOMP} \citep{chu2022transphla}: The source code in the \href{https://github.com/a96123155/TransPHLA-AOMP}{authors' repository} is GPL-3.0 licensed. The pMHC comparison also uses the authors' released checkpoint (\href{https://github.com/a96123155/TransPHLA-AOMP/blob/3ed2260292934170757507a71e645d0bcadfc44b/TransPHLA-AOMP/pHLAIformer.pkl}{\texttt{pHLAIformer.pkl}}).
    \item \textbf{pMTnet} \citep{lu2021pMTnet}: The \href{https://github.com/tianshilu/pMTnet}{source repository} lists GPL-2.0 for its code. The released-checkpoint pMT comparison uses the authors' \href{https://github.com/tianshilu/pMTnet/blob/f9244234be1bc80a310a3dc04c02093b58e71b5e/library/h5_file/HLA_antigen_encoder_60.h5}{pMHC} and \href{https://github.com/tianshilu/pMTnet/blob/f9244234be1bc80a310a3dc04c02093b58e71b5e/library/h5_file/TCR_encoder_30.h5}{TCR} encoders and \href{https://github.com/tianshilu/pMTnet/blob/f9244234be1bc80a310a3dc04c02093b58e71b5e/library/h5_file/weights.h5}{prediction head}; retrained configurations are described in \Cref{exp:abl}. The NetMHCpan baseline in Figure~\ref{fig:ModelPerformance} uses the released encoder and a retrained linear classifier.
    \item \textbf{PISTE} \citep{feng2024sliding}: At \href{https://github.com/jychen01/PISTE/tree/ca782c34c48d5471b47f420c21bd04d2b110bb2f}{repository revision ca782c3}, code is MIT licensed and data are CC0 1.0 licensed. The comparison includes a retrained model and the authors' released \href{https://github.com/jychen01/PISTE/blob/ca782c34c48d5471b47f420c21bd04d2b110bb2f/code/checkpoints/random/exp0.pkl}{random checkpoint}.
    \item \textbf{ERGO II} \citep{springer2021ergo2}: The \href{https://github.com/IdoSpringer/ERGO-II}{authors' code repository} is MIT licensed. The comparison includes both a retrained model and the authors' \href{https://github.com/IdoSpringer/ERGO-II/blob/85d320ab03ade33460be9a81ea3a51b8d37cd998/Models/version_1veajht/checkpoints/_ckpt_epoch_10.ckpt}{released checkpoint}.
    \item \textbf{ESM-2}: The \href{https://github.com/facebookresearch/esm}{source code} and the \href{https://huggingface.co/facebook/esm2_t6_8M_UR50D}{8M model card} and \href{https://huggingface.co/facebook/esm2_t12_35M_UR50D}{35M model card} list MIT. Our MLP baseline uses ESM-2 8M sequence representations.
    \item \textbf{ANN 4.0}: Predictions were obtained through the \href{https://tools.iedb.org/mhci/}{IEDB MHC-I binding prediction service}.
    \item \textbf{PickPocket 1.1}: Predictions were obtained through the \href{https://services.healthtech.dtu.dk/services/PickPocket-1.1/}{DTU Health Tech web service}.
\end{itemize}

The listed licenses apply to the linked source repositories. Separately distributed checkpoints and datasets retain their respective distribution terms.

\section{Detailed Numerical and Statistical Results}
\label{app:additional_experiments}

\subsection{Baseline results}
\label{app:baseline-results}

Table~\ref{tab:baseline_current} lists the five testing metrics underlying the baseline figures, separating retraining variation from fixed-model predictions. For completeness, it also includes the pMT results reported in Table~\ref{tab:pmt_baselines_main}.

On the pMHC testing dataset, ANN and PickPocket retain 141,162 and 167,082 examples, respectively, after input validity checks. The other models use all 171,438 examples. For class-balanced evaluation, the pMT metrics assign weight 10 to positives and 1 to negatives, while all pMHC metrics use the native class proportions of the evaluated examples without reweighting. The NetMHCpan baseline uses released frozen features and a retrained linear head. The MLP uses ESM-2 8M with concatenated sequence representations.
\par

\begin{table*}[htbp]
\centering
\caption{\textbf{Comprehensive baseline performance across both evaluation tasks (\%).} Retrained scores represent the mean $\pm$ sample SD over five independent runs; released predictors yield a single deterministic score. Metric conventions and pMHC test coverage are described above. See Section~\ref{exp:bas} for pMT test coverage.}

\label{tab:baseline_current}
\footnotesize
\setlength{\tabcolsep}{3pt}
\begin{tabular*}{\textwidth}{@{\extracolsep{\fill}}llccccc}
\toprule
\textbf{Model} & \textbf{Version} & \textbf{ROC AUC} & \textbf{PR AUC} & \textbf{MCC} & \textbf{ACC} & \textbf{F1} \\
\midrule
\multicolumn{7}{c}{\textbf{pMHC testing}} \\
\addlinespace[2pt]
\textbf{Fusion-pM} & Retrained & $98.40\pm0.01$ & $98.33\pm0.02$ & $88.17\pm0.10$ & $94.08\pm0.05$ & $94.10\pm0.05$ \\
NetMHCpan & Retrained & $82.64\pm0.12$ & $79.58\pm0.16$ & $53.63\pm0.16$ & $76.17\pm0.09$ & $78.57\pm0.08$ \\
TransPHLA & Retrained & $96.82\pm0.06$ & $96.69\pm0.07$ & $81.90\pm0.18$ & $90.95\pm0.09$ & $91.00\pm0.10$ \\
TransPHLA & Checkpoint & $97.89$ & $97.65$ & $86.17$ & $93.08$ & $93.09$ \\
ANN & Official predictor & $93.19$ & $93.99$ & $61.38$ & $78.01$ & $72.44$ \\
PickPocket & Official predictor & $93.01$ & $92.98$ & $49.68$ & $70.47$ & $58.94$ \\
\midrule
\multicolumn{7}{c}{\textbf{pMT testing}} \\
\addlinespace[2pt]
\textbf{Fusion-pMT} & Retrained & $88.48\pm2.03$ & $90.46\pm1.30$ & $62.69\pm3.31$ & $80.76\pm1.92$ & $78.70\pm2.70$ \\
PISTE & Retrained & $85.72\pm1.48$ & $86.00\pm1.78$ & $54.19\pm6.38$ & $76.37\pm3.58$ & $73.91\pm5.13$ \\
PISTE & Checkpoint & $73.61$ & $68.85$ & $33.34$ & $66.55$ & $65.40$ \\
pMTnet & Retrained & $79.85\pm2.02$ & $82.54\pm1.53$ & $49.98\pm3.24$ & $73.86\pm1.88$ & $69.25\pm2.99$ \\
pMTnet & Checkpoint & $68.02$ & $73.15$ & $32.88$ & $61.02$ & $38.02$ \\
ERGO II & Retrained & $48.93\pm6.78$ & $48.62\pm4.41$ & $-3.91\pm6.13$ & $48.27\pm2.74$ & $27.36\pm27.33$ \\
ERGO II & Checkpoint & $52.55$ & $54.20$ & $1.08$ & $50.39$ & $45.15$ \\
MLP & Retrained & $83.05\pm1.51$ & $83.97\pm1.28$ & $46.36\pm3.43$ & $70.43\pm2.29$ & $61.16\pm4.53$ \\
\bottomrule
\end{tabular*}
\end{table*}

\subsection{Ablation Implementation Details}
\label{app:ablation-results}
Table~\ref{tab:ablation_current} in the main text reports the complete-model reference, component ablations, and progressive comparison.

In the \textit{w/o Class Attention} condition, the replacement Stage~2 readout contains 50,372 parameters, compared with 50,304 in the original class-attention readout. These counts refer only to the readout modules, not the complete models.

\subsection{Seed-level statistical comparisons}
\label{app:seed-comparisons}

To quantify training-seed variation on the fixed testing sets, we conduct exploratory one-sided $t$-tests evaluating whether Fusion-pM or Fusion-pMT achieves a higher expected score than each comparator (\Cref{tab:model_comparison}). Specifically, we apply paired $t$-tests across matched training seeds, Welch's $t$-tests for independently seeded runs, and one-sample $t$-tests against fixed released predictors (which evaluate the five complete-model scores relative to the deterministic baseline on matching test examples). Reported $p$-values are unadjusted for multiple comparisons.

In \Cref{tab:model_comparison}, dashes denote comparators evaluated on different retained test subsets (ANN, PickPocket, both PISTE versions, and the official ERGO II checkpoint), for which matched complete-model subset scores are not computed.

\begin{table*}[htbp]
\centering
\caption{\textbf{Exploratory one-sided $t$-test $p$-values across evaluation tasks and ablations.} \textit{Paired}, \textit{Welch}, and \textit{Fixed} denote the paired, two-sample Welch, and one-sample $t$-tests defined in Appendix~\ref{app:seed-comparisons} across five independent runs. Shared baseline comparators appearing in multiple panels yield identical values, and dashes (--) indicate models evaluated on different retained subsets.}
\label{tab:model_comparison}
\footnotesize
\setlength{\tabcolsep}{3pt}
\resizebox{\textwidth}{!}{\begin{tabular}{llccccc}
\toprule
\textbf{Comparator} & \textbf{Test} & \textbf{ROC AUC} & \textbf{PR AUC} & \textbf{MCC} & \textbf{ACC} & \textbf{F1} \\
\midrule
\multicolumn{7}{l}{\textbf{\Cref{fig:ModelPerformance}}} \\
NetMHCpan, retrained & Welch & $3.8\times10^{-10}$ & $4.1\times10^{-10}$ & $4.6\times10^{-16}$ & $2.4\times10^{-15}$ & $3.0\times10^{-16}$ \\
TransPHLA, retrained & Welch & $1.2\times10^{-7}$ & $1.7\times10^{-7}$ & $2.1\times10^{-10}$ & $1.6\times10^{-10}$ & $7.0\times10^{-10}$ \\
TransPHLA, checkpoint & Fixed & $1.7\times10^{-8}$ & $3.9\times10^{-8}$ & $7.4\times10^{-7}$ & $7.5\times10^{-7}$ & $8.3\times10^{-7}$ \\
ANN, official & -- & -- & -- & -- & -- & -- \\
PickPocket, official & -- & -- & -- & -- & -- & -- \\
\midrule
\multicolumn{7}{l}{\textbf{\Cref{fig:Model Performance}}} \\
PISTE, retrained & -- & -- & -- & -- & -- & -- \\
PISTE, checkpoint & -- & -- & -- & -- & -- & -- \\
pMTnet, retrained & Paired & $0.0016$ & $4.0\times10^{-4}$ & $0.0023$ & $0.0030$ & $0.0044$ \\
pMTnet, checkpoint & Fixed & $1.1\times10^{-5}$ & $3.7\times10^{-6}$ & $1.8\times10^{-5}$ & $1.1\times10^{-5}$ & $2.3\times10^{-6}$ \\
ERGO II, retrained & Welch & $4.3\times10^{-5}$ & $4.7\times10^{-6}$ & $2.6\times10^{-7}$ & $4.2\times10^{-8}$ & $0.0067$ \\
ERGO II, checkpoint & -- & -- & -- & -- & -- & -- \\
MLP, retrained & Welch & $8.4\times10^{-4}$ & $2.3\times10^{-5}$ & $3.0\times10^{-5}$ & $3.3\times10^{-5}$ & $1.0\times10^{-4}$ \\
\midrule
\multicolumn{7}{l}{\textbf{\Cref{fig:Model Performance2}(a)}} \\
w/o Shared Projection & Paired & $0.1241$ & $0.0439$ & $0.0199$ & $0.0249$ & $0.0333$ \\
w/o Cross-Attention & Paired & $0.0598$ & $0.0328$ & $0.0521$ & $0.0698$ & $0.1086$ \\
w/o Class Attention & Paired & $0.0557$ & $0.0383$ & $0.0396$ & $0.0341$ & $0.0294$ \\
w/o Pretraining & Paired & $0.0226$ & $0.0182$ & $0.0525$ & $0.0536$ & $0.0555$ \\
\midrule
\multicolumn{7}{l}{\textbf{\Cref{fig:Model Performance2}(b)}} \\
pMTnet pMHC Representation, retrained & Paired & $0.1653$ & $0.1015$ & $0.0939$ & $0.1539$ & $0.3297$ \\
pMTnet pMT Representation, retrained & Paired & $0.0023$ & $8.7\times10^{-4}$ & $0.0029$ & $0.0032$ & $0.0035$ \\
pMTnet, retrained & Paired & $0.0016$ & $4.0\times10^{-4}$ & $0.0023$ & $0.0030$ & $0.0044$ \\
pMTnet, checkpoint & Fixed & $1.1\times10^{-5}$ & $3.7\times10^{-6}$ & $1.8\times10^{-5}$ & $1.1\times10^{-5}$ & $2.3\times10^{-6}$ \\
\bottomrule
\end{tabular}}
\end{table*}

\subsection{Supplementary Comparisons with ESM-2 Representations}
\label{app:esm2_comparison}
\label{exp:extension}

We evaluate frozen and fine-tuned ESM-2 representations
\citep{lin2023evolutionary} on the pMT test set.
Table~\ref{tab:esm2_comparison} includes the original Fusion-pMT as a
complete-model reference. The \texttt{Concat} baseline applies masked mean
pooling to each molecular sequence, concatenates the pooled vectors, and
feeds them to an MLP classifier. The fine-tuned ESM-2 8M \texttt{Concat}
configuration is the MLP baseline in Figure~\ref{fig:Model Performance}.

In the \texttt{ESM-2 + Fusion-pMT} configurations, a shared ESM-2 backbone
(8M or 35M) produces residue embeddings that are linearly projected to
64 dimensions and passed to the full two-stage Fusion-pMT architecture.
In the \texttt{Frozen} setting, the ESM-2 weights remain fixed while the
downstream modules are trained. In the \texttt{Fine-tune} setting, training
also updates the ESM-2 weights. These configurations train for eight epochs
per stage. We select the checkpoint at each stage by validation PR AUC,
transfer the selected Stage~1 checkpoint to Stage~2, and select the
classification threshold by validation MCC.

All configurations are evaluated on the same 6,809 test examples using
weights of 10 for positives and 1 for negatives. The original Fusion-pMT
and the \texttt{ESM-2 + Fusion-pMT} configurations use seeds 4001, 4003,
4007, 4013, and 4019. The \texttt{Concat} configurations use seeds 42--46.
Threshold-dependent metrics use decision thresholds selected on the
validation set.

\begin{table*}[htbp]
\centering
\caption{\textbf{Fusion-pMT and ESM-2 pipeline performance (\%).} Values are mean $\pm$ sample SD across five runs. Pipeline definitions and evaluation settings appear above. Bold indicates the higher mean PR AUC within each ESM-2 backbone and training setting.}
\label{tab:esm2_comparison}
\footnotesize
\setlength{\tabcolsep}{3pt}
\resizebox{\textwidth}{!}{\begin{tabular}{lllccccc}
\toprule
\textbf{Backbone} & \textbf{Fusion} & \textbf{Training} & \textbf{ROC AUC} & \textbf{PR AUC} & \textbf{MCC} & \textbf{ACC} & \textbf{F1} \\
\midrule
\multicolumn{3}{l}{Fusion-pMT (original)} & $88.48\pm2.03$ & $90.46\pm1.30$ & $62.69\pm3.31$ & $80.76\pm1.92$ & $78.70\pm2.70$ \\
\midrule
\multirow{2}{*}{ESM-2 8M} & Concat & \multirow{2}{*}{Frozen} & $70.80\pm0.74$ & $70.15\pm0.54$ & $12.17\pm3.96$ & $52.61\pm0.65$ & $67.17\pm0.75$ \\
 & Fusion-pMT &  & $85.40\pm3.19$ & {\boldmath $86.86\pm2.62$} & $50.71\pm10.90$ & $74.38\pm6.70$ & $77.02\pm4.02$ \\
\cmidrule(lr){1-8}
\multirow{2}{*}{ESM-2 8M} & Concat & \multirow{2}{*}{Fine-tune} & $83.05\pm1.51$ & $83.97\pm1.28$ & $46.36\pm3.43$ & $70.43\pm2.29$ & $61.16\pm4.53$ \\
 & Fusion-pMT &  & $86.64\pm1.48$ & {\boldmath $86.51\pm2.11$} & $56.89\pm4.98$ & $78.37\pm2.54$ & $77.63\pm3.17$ \\
\midrule
\multirow{2}{*}{ESM-2 35M} & Concat & \multirow{2}{*}{Frozen} & $75.52\pm0.61$ & $75.82\pm1.27$ & $19.06\pm3.05$ & $54.91\pm1.36$ & $68.44\pm0.55$ \\
 & Fusion-pMT &  & $86.94\pm1.47$ & {\boldmath $87.87\pm1.30$} & $50.14\pm4.87$ & $73.84\pm3.14$ & $77.31\pm1.67$ \\
\cmidrule(lr){1-8}
\multirow{2}{*}{ESM-2 35M} & Concat & \multirow{2}{*}{Fine-tune} & $78.51\pm6.45$ & $79.79\pm4.93$ & $37.99\pm9.45$ & $67.70\pm4.81$ & $60.60\pm7.58$ \\
 & Fusion-pMT &  & $81.25\pm3.85$ & {\boldmath $82.49\pm3.16$} & $46.24\pm6.05$ & $72.73\pm3.39$ & $70.11\pm5.74$ \\
\bottomrule
\end{tabular}}
\end{table*}

\paragraph{Results and key observations.}
Across all four ESM-2 backbone and training settings, the Fusion-pMT
pipeline achieved higher mean scores than the corresponding
\texttt{Concat} baseline on all five metrics
(Table~\ref{tab:esm2_comparison}). Beyond the large gains under frozen
backbones reported in Section~\ref{exp:abl}, Fusion-pMT maintained clear advantages
after end-to-end fine-tuning, exceeding \texttt{Concat} by 2.54 and 2.70
percentage points in mean PR AUC for fine-tuned ESM-2 8M and 35M,
respectively.

For ESM-2 35M, fine-tuning increased mean PR AUC for \texttt{Concat}
from 75.82\% to 79.79\% and reduced it for Fusion-pMT from 87.87\% to
82.49\%. The sample standard deviation of PR AUC increased across runs
for both pipelines. Frozen ESM-2 35M with Fusion-pMT achieved the highest
mean PR AUC among the ESM-2 configurations, yet remained below the
original Fusion-pMT (90.46\% under its standard training schedule),
indicating that competitive performance can be achieved without large
pretrained backbones. These comparisons reflect the complete pipelines,
including their fusion and training procedures.

This discrepancy is biologically and computationally grounded: (1) MHC pseudosequences consist of discontinuous contact residues and TCR CDR3 loops arise from somatic V(D)J recombination rather than natural phylogenetic evolution, partially diverging from the natural evolutionary masked-language priors captured by ESM-2; and (2) fine-tuning high-capacity representations on specialized binding pairs increases susceptibility to overfitting, as evidenced by higher variance across runs. Fusion-pMT's compact task-tailored sequence projection avoids these issues while retaining sample efficiency.

\end{document}